\documentclass[twocolumn]{aastex63}

\usepackage{CJK}
\usepackage{float}
\usepackage{ulem}

\received{XX XX, 2020}
\revised{XX XX, 2020}
\accepted{XXX}
\submitjournal{XXX}

\shorttitle{2018cuf}
\shortauthors{Yi-Ze Dong et al.}

\graphicspath{{./}{figures/}}
\begin{document}
\title{Supernova 2018cuf: A Type IIP supernova with a slow fall from plateau}
\begin{CJK*}{UTF8}{gbsn}
\author[0000-0002-7937-6371]{Yize Dong (董一泽)}
\affiliation{Department of Physics and Astronomy, University of California, 1 Shields Avenue, Davis, CA 95616-5270, USA}
\author[0000-0001-8818-0795]{S. Valenti}
\affiliation{Department of Physics and Astronomy, University of California, 1 Shields Avenue, Davis, CA 95616-5270, USA}
\author[0000-0002-4924-444X]{K. A. Bostroem}
\affiliation{Department of Physics and Astronomy, University of California, 1 Shields Avenue, Davis, CA 95616-5270, USA}
\author[0000-0003-4102-380X]{D. J. Sand}
\affiliation{Steward Observatory, University of Arizona, 933 North Cherry Avenue, Rm. N204, Tucson, AZ 85721-0065, USA}
\author[0000-0003-0123-0062]{Jennifer E. Andrews}
\affiliation{Steward Observatory, University of Arizona, 933 North Cherry Avenue, Rm. N204, Tucson, AZ 85721-0065, USA}
\author[0000-0002-1296-6887]{L. Galbany}
\affiliation{Departamento de F\'isica Te\'orica y del Cosmos, Universidad de Granada, E-18071 Granada, Spain.}
\author[0000-0001-8738-6011]{Saurabh W. Jha}
\affiliation{Department of Physics and Astronomy, Rutgers, the State University of New Jersey, 136 Frelinghuysen Road, Piscataway, NJ 08854, USA}
\author{Youssef Eweis}
\affiliation{Department of Physics and Astronomy, Rutgers, the State University of New Jersey, 136 Frelinghuysen Road, Piscataway, NJ 08854, USA}
\affiliation{Department of Physics and Astronomy, Iowa State University, Ames, IA 50011, USA}
\author{Lindsey Kwok}
\affiliation{Department of Physics and Astronomy, Rutgers, the State University of New Jersey, 136 Frelinghuysen Road, Piscataway, NJ 08854, USA}
\author[0000-0003-1039-2928]{E. Y. Hsiao}
\affiliation{Department of Physics, Florida State University, 77 Chieftan Way, Tallahassee, FL 32306, USA}
\author[0000-0002-2806-5821]{Scott Davis}
\affiliation{Department of Physics, Florida State University, 77 Chieftan Way, Tallahassee, FL 32306, USA}
\author[0000-0001-6272-5507]{Peter J. Brown}
\affiliation{Department of Physics and Astronomy, Texas A\&M University, 4242 TAMU, College Station, TX 77843, US}
\author{H. Kuncarayakti}
\affiliation{Tuorla Observatory, Department of Physics and Astronomy, University of Turku, Turku FI-20014, Finland}
\affiliation{Finnish Centre for Astronomy with ESO (FINCA), University of Turku, Turku FI-20014, Finland}
\author[0000-0003-2611-7269]{Keiichi Maeda}
\affiliation{Department of Astronomy, Graduate School of Science, Kyoto University, Sakyo-ku, Kyoto 606-8502, Japan}
\affiliation{Kavli Institute for the Physics and Mathematics of the Universe (WPI), The University of Tokyo, 5-1-5 Kashiwanoha, Kashiwa, Chiba 277-8583, Japan}
\author{Jeonghee Rho}
\affiliation{SETI Institute, 189 N. Bernardo Ave, Mountain View, CA 94043}
\author[0000-0002-1546-9763]{R. C. Amaro}
\affiliation{Steward Observatory, University of Arizona, 933 North Cherry Avenue, Rm. N204, Tucson, AZ 85721-0065, USA}
\author[0000-0003-0227-3451]{J. P. Anderson}
\affiliation{European Southern Observatory, Alonso de C\'ordova 3107, Vitacura, Casilla 190001, Santiago, Chile}
\author[0000-0001-7090-4898]{Iair Arcavi}
\affiliation{The School of Physics and Astronomy, Tel Aviv University, Tel Aviv 69978, Israel}
\affiliation{CIFAR Azrieli Global Scholars program, CIFAR, Toronto, Canada}
\author{Jamison Burke}
\affiliation{Department of Physics, University of California, Santa Barbara, CA 93106-9530}
\affiliation{Las Cumbres Observatory, 6740 Cortona Drive, Suite 102, Goleta, CA 93117-5575, USA}
\author[0000-0001-6191-7160]{Raya Dastidar}
\affiliation{Aryabhatta Research Institute of observational sciences, Manora Peak, Nainital 263 001, India}
\affiliation{Department of Physics $\&$ Astrophysics, University of Delhi, Delhi-110 007}
\author{Gast\'on Folatelli}
\affiliation{Kavli Institute for the Physics and Mathematics of the Universe (WPI), The University of Tokyo, 5-1-5 Kashiwanoha, Kashiwa, Chiba 277-8583, Japan}
\affiliation{Facultad de Ciencias Astron\'omicas y Geof\'isicas, Universidad Nacional de La Plata, Paseo del Bosque S/N, B1900FWA La Plata, Argentina}
\affiliation{Instituto de Astrof\'isica de La Plata (IALP), CONICET, Argentina}
\author{Joshua Haislip}
\affiliation{Department of Physics and Astronomy, University of North Carolina at Chapel Hill, Chapel Hill, NC 27599, USA}
\author[0000-0002-1125-9187]{Daichi Hiramatsu}
\affiliation{Department of Physics, University of California, Santa Barbara, CA 93106-9530}
\affiliation{Las Cumbres Observatory, 6740 Cortona Drive, Suite 102, Goleta, CA 93117-5575, USA}
\author[0000-0002-0832-2974]{Griffin Hosseinzadeh}
\affiliation{Center for Astrophysics \textbar{} Harvard \& Smithsonian, 60 Garden Street, Cambridge, MA 02138-1516, USA}
\author[0000-0003-4253-656X]{D. Andrew Howell}
\affiliation{Department of Physics, University of California, Santa Barbara, CA 93106-9530}
\affiliation{Las Cumbres Observatory, 6740 Cortona Drive, Suite 102, Goleta, CA 93117-5575, USA}
\author[0000-0001-5754-4007]{J. Jencson}
\affiliation{Steward Observatory, University of Arizona, 933 North Cherry Avenue, Rm. N204, Tucson, AZ 85721-0065, USA}
\author[0000-0003-3642-5484]{Vladimir Kouprianov}
\affiliation{Department of Physics and Astronomy, University of North Carolina at Chapel Hill, Chapel Hill, NC 27599, USA}
\author[0000-0001-9589-3793]{M. Lundquist}
\affiliation{Steward Observatory, University of Arizona, 933 North Cherry Avenue, Rm. N204, Tucson, AZ 85721-0065, USA}
\author[0000-0002-3464-0642]{J. D. Lyman}
\affiliation{Department of Physics, University of Warwick, Coventry CV4 7AL, UK}
\author{Curtis McCully}
\affiliation{Department of Physics, University of California, Santa Barbara, CA 93106-9530}
\affiliation{Las Cumbres Observatory, 6740 Cortona Drive, Suite 102, Goleta, CA 93117-5575, USA}
\author[0000-0003-1637-267X]{Kuntal Misra}
\affiliation{Aryabhatta Research Institute of observational sciences, Manora Peak, Nainital 263 001, India}
\author[0000-0002-5060-3673]{Daniel E. Reichart}
\affiliation{Department of Physics and Astronomy, University of North Carolina at Chapel Hill, Chapel Hill, NC 27599, USA}
\author{S. F. S\'anchez}
\affiliation{Instituto de Astronom\'ia, Universidad Nacional Aut\'onoma de M\'exico Circuito Exterior, Ciudad Universitaria, Ciudad de M\'exico 04510, Mexico}
\author{Nathan Smith}
\affiliation{Steward Observatory, University of Arizona, 933 North Cherry Avenue, Rm. N204, Tucson, AZ 85721-0065, USA}
\author{Xiaofeng Wang}
\affiliation{Physics Department and Tsinghua Center for Astrophysics (THCA), Tsinghua University, Beijing, 100084, People’s Republic of China}
\author{Lingzhi Wang}
\affiliation{CAS Key Laboratory of Optical Astronomy, National Astronomical Observatories, Chinese Academy of Sciences, Beijing 100101, China}
\affiliation{Chinese Academy of Sciences South America Center for Astronomy, National Astronomical Observatories, CAS, Beijing 100101, China}
\author[0000-0003-2732-4956]{S. Wyatt}
\affiliation{Steward Observatory, University of Arizona, 933 North Cherry Avenue, Rm. N204, Tucson, AZ 85721-0065, USA}


\begin{abstract}

We present multi-band photometry and spectroscopy of SN~2018cuf, a Type IIP (``P" for plateau) supernova (SN) discovered by the Distance Less Than 40 Mpc survey (DLT40) within 24 hours of explosion.
SN~2018cuf appears to be a typical Type IIP SN, with an absolute $V$-band magnitude of $-$16.73 $\pm$ 0.32 at maximum and a decline rate of 0.21 $\pm$ 0.05 mag/50d during the plateau phase. 
The distance of the object is constrained to be 41.8 $\pm$ 5.7 Mpc by using the expanding photosphere method.
We use spectroscopic and photometric observations from the first year after the explosion to constrain the progenitor of SN~2018cuf using both hydrodynamic light curve modelling and late-time spectroscopic modelling.
The progenitor of SN~2018cuf was most likely a red supergiant of about 14.5 $\rm M_{\odot}$ that produced 0.04 $\pm$ 0.01 $\rm M_{\odot}$ $\rm ^{56}Ni$ during the explosion. We also found $\sim$ 0.07 $\rm M_{\odot}$ of circumstellar material (CSM) around the progenitor is needed to fit the early light curves, where the CSM may originate from pre-supernova outbursts.
During the plateau phase, high velocity features at $\rm \sim 11000\ km~s^{-1}$ are detected both in the optical and near-infrared spectra, supporting the possibility that the ejecta were interacting with some CSM. A very shallow slope during the post-plateau phase is also observed and it is likely due to a low degree of nickel mixing or the relatively high nickel mass in the SN.
\end{abstract}

\keywords{supernovae: individual (SN 2018cuf)}

\section{Introduction} 
\label{sec:intro}
Type II supernovae (SNe), the most common type of core-collapse supernova (CCSN), originate from the collapse of stars more massive than $\sim$8 $\rm M_{\odot}$. In the Type IIP subclass, the SN experiences a period of nearly constant luminosity for $\sim$ 2--3 months after maximum as the hydrogen envelope recombines. This is then followed by a rapid drop from the plateau where the light curve becomes dominated by radioactive decay and the SN enters the nebular phase. 

From pre-explosion imaging at the location of the explosions, the progenitors of Type IIP SNe have been mostly attributed to red supergiants (RSGs) with initial masses of $\sim$ 8--17 $\rm M_{\odot}$ \citep{VanDyk03,Smartt09,Smartt2015}. However, evolutionary codes predict that the progenitors of Type IIP SNe can have masses up to 30 $\rm M_{\odot}$ \citep[e.g.][]{Heger2003,Ekstrom2012}. This discrepancy between observations and theory has been dubbed the  ``red supergiant (RSG) problem.'' This problem has been discussed by many authors \citep[e.g.][]{Walmswell2012,Kochanek2012,Horiuchi2014,Davies2018,Davies2020}, and remains an open question. An alternative method that is widely used to estimate the progenitor masses of Type II SNe is hydrodynamic modelling of SN light curves \citep[e.g.][]{Utrobin2015,Utrobin2017,Morozova2017,Morozova2018,Paxton2018,Goldberg2019,Martinez2019}. Through comparing observed light curves with model light curves, many progenitor properties, such as mass, radius and explosion energy, could be determined. Another approach to estimate the progenitor mass is nebular spectral modelling \citep{Jerkstrand2012,Jerkstrand2014}. Here the structure and composition of the ejecta can be constrained, and the intensity of the [\ion{O}{1}]~$\lambda\lambda$6300,6363 doublet can be used to derive the progenitor mass. 

These various methods sometimes do not predict a consistent progenitor mass for a given SN, so continued observational and theoretical work is necessary for these different techniques to converge \citep{Jerkstrand2014,Morozova2018,Davies2018}. The progenitor mass distribution inferred from hydrodynamic modelling is generally higher than the observed mass range from direct imaging, mitigating the RSG problem \citep[although see \cite{Martinez2020arXiv where consistent masses are obtained between hydrodynamical modelling and other methods}]{Morozova2018}. On the other hand, \cite{Jerkstrand2014} found that, from nebular spectral modelling, there is no evidence yet that the progenitor of an observed Type II SN is more massive than 20 $\rm M_{\odot}$, supporting the presence of the RSG problem. However, some recent SN studies have found more massive progenitors based on nebular spectral modelling \citep{anderson2018,bose2020}. It is important to note that the sample of SNe that have been studied by these two modelling techniques is small. Increasing the sample size is necessary to fully examine the existence of the RSG problem.

For this purpose, observations both in the first few days after explosion and during the nebular phase ($\sim$300-500 days after explosion) are required. Unfortunately rapid discovery and follow-up of SNe is still rare, and often Type IIP SNe are not followed out to the nebular phase when larger telescopes are needed. Thankfully, modern SN surveys such as the All Sky Automated Survey for SNe (ASAS-SN, \citealt{Shappee14,Kochanek2017}), the Zwicky Transient Facility (ZTF, \citealt{Bellm2019}), the Asteroid Terrestrial-Impact Last Alert System (ATLAS, \citealt{Tonry2011,ATLAS2}), and the Distance Less Than 40 Mpc survey (DLT40, \citealt{Tartaglia2018}) are now able to discover SNe within hours of explosion and use dedicated facilities for follow-up, such as the Las Cumbres Observatory \citep[LCOGT]{Brown2013}. The very early light curves of core-collapse SNe provided by these surveys can be used to constrain the progenitor radius (and potentially the envelope structure), ejected mass, and kinetic energy of the explosion \citep[e.g.][for selected theoretical and observational results]{Rabinak2011,Sapir2017,Bersten18, Arcavi17,Piro17}.

In this paper, we present optical and infrared photometry and spectroscopy of SN~2018cuf, a Type II SN discovered within 30 hours of explosion by the DLT40 survey and densely monitored within the Global SN Project (GSP)\footnote{GSP is a Key Project at Las Cumbres Observatory} for over 340 days. 
This paper is organized as follows: the observations of SN~2018cuf are presented in Section \ref{sec:obs}, while the reddening and host galaxy properties are presented in Section \ref{sec:reddening}. Further observational properties, such as the distance and explosion epoch, are constrained in Section \ref{sec:obs_properties}. In Section \ref{sec:photo_evol} we analyze the light curves and in Section \ref{sec:spec_evol} the spectroscopic evolution is described. We constrain the nickel mass and progenitor mass using our extensive observational data set in Section \ref{sec:progenitor}, and finally we present our conclusions in Section \ref{sec:conclusion}.

\begin{figure}
\includegraphics[width=1.15\linewidth]{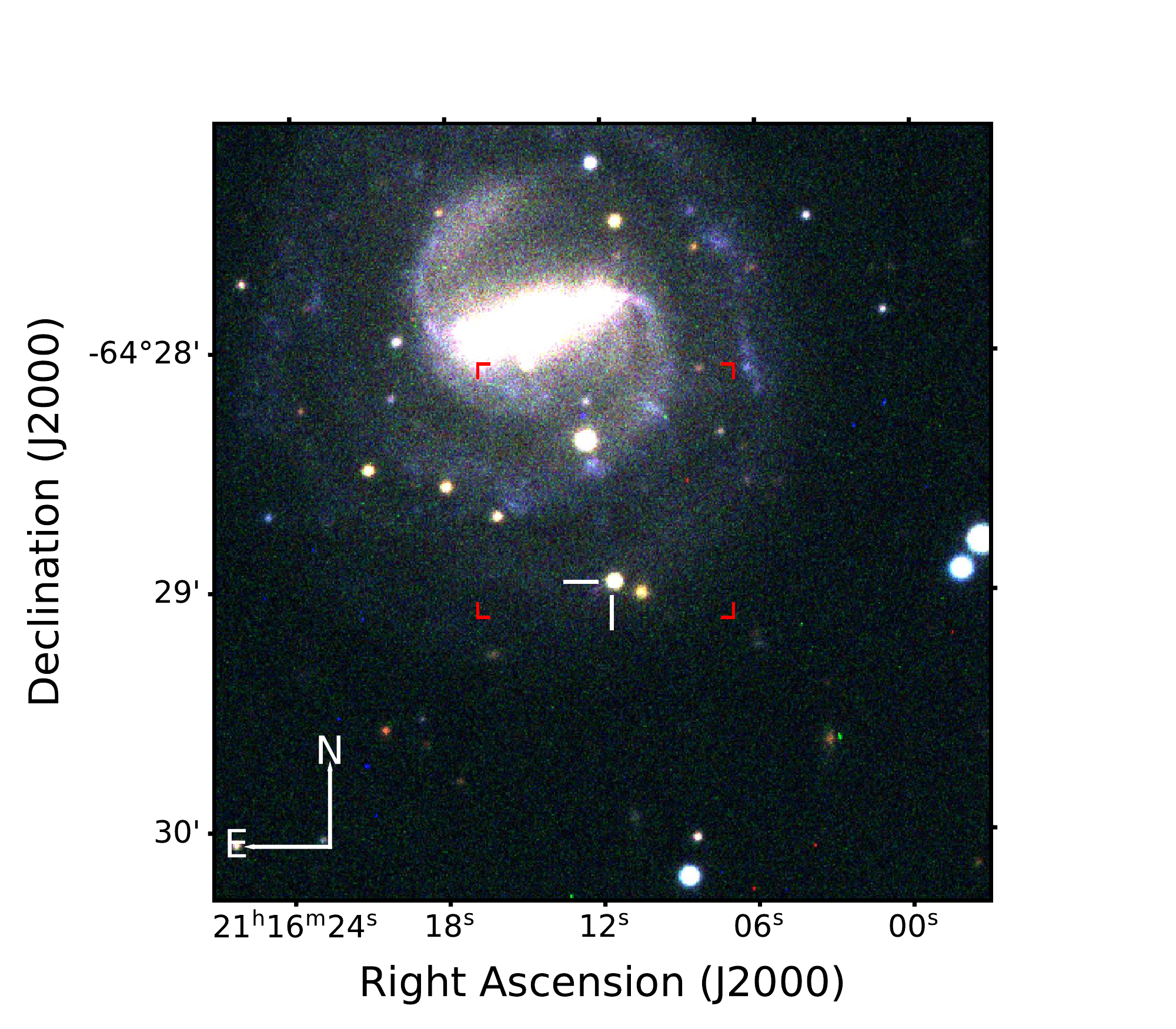}
\caption{RGB image of SN~2018cuf (indicated by white tick marks) in IC~5092 obtained with the Las Cumbres Observatory on 2018 September 17.  The red markers delineate the MUSE field of view, as described in Section~3.2.  \label{fig:sn_image}}
\end{figure}

\begin{figure*}
\includegraphics[width=1\linewidth]{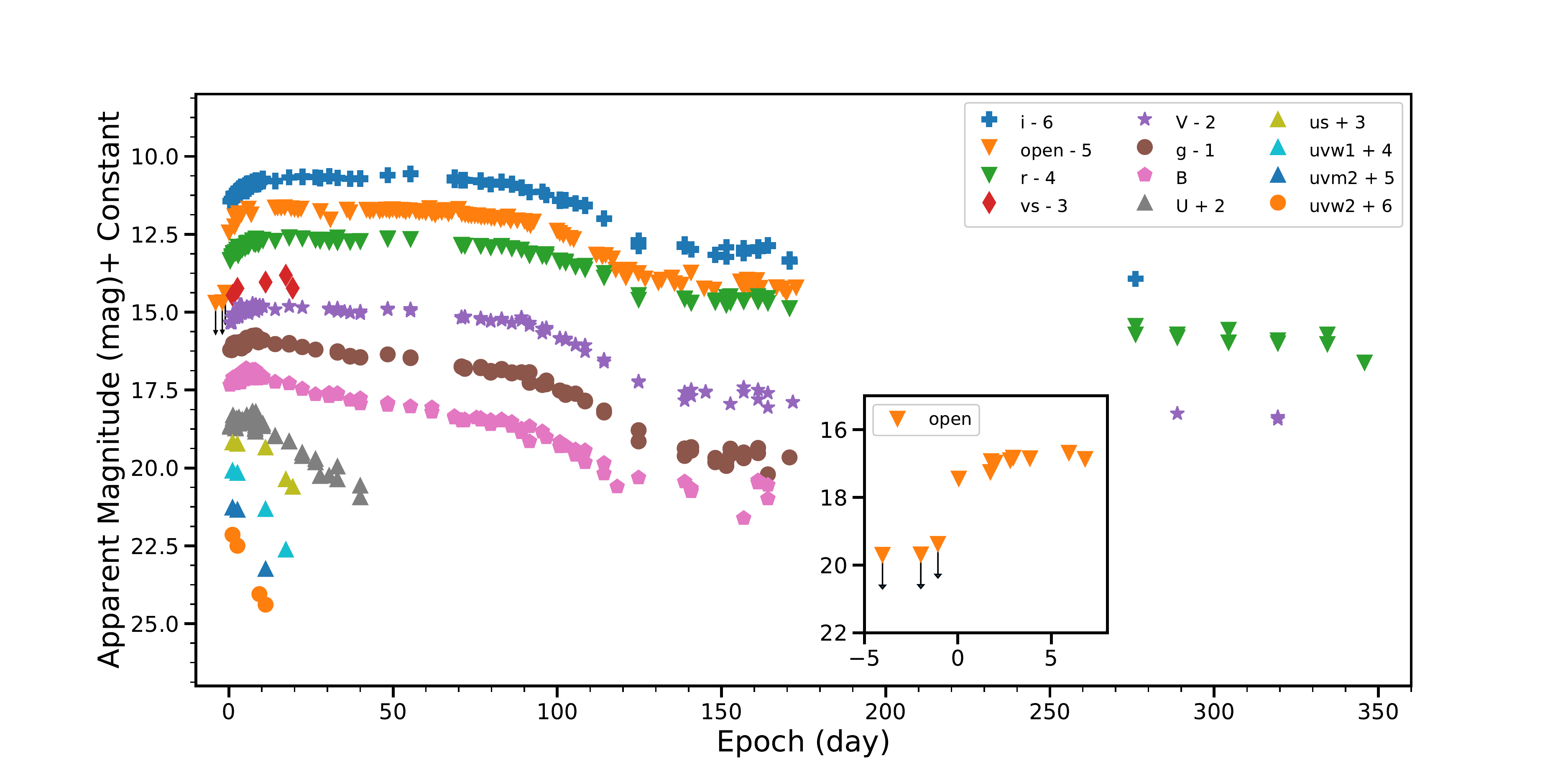}
\caption{Multi-band light curves for SN~2018cuf with respect to the epoch of explosion. 
An $Open$ filter is used by the PROMPT5 0.4 m telescope and is calibrated to the $r$-band. The insert is a zoom on the $Open$ filter illustrating the DLT40 detection limit $\sim$1 day before discovery. \label{fig:photometry}}
\end{figure*}

\section{Observations} \label{sec:obs}
SN~2018cuf was discovered at RA(2000) $=$ 21$^h$16$^m$11$^s$.58, Dec(2000) $=-64\degr 28\arcmin 57\farcs 30$ in the nearby SBc galaxy IC~5092 (see Figure \ref{fig:sn_image}) on 2018 June 23 (\citealt{Valenti2018}, JD 2458292.86093, $r$ = 17.4), during the course of the DLT40 SN search \citep{Tartaglia2018}, utilizing the 0.4-m \rm{PROMPT5} telescope \citep{Reichart2005} at the Cerro Tololo Inter-American Observatory (CTIO). 
A non-detection $\sim$24 hours earlier (JD 2458291.74456; $r$ $\lesssim$ 19.4) strongly constrains the explosion epoch (see Figure \ref{fig:photometry}). The 1-day cadence of the DLT40 SN search is designed to discover $\sim$ 10 nearby SNe ($<$ 40 Mpc) per year within 24 hours of explosion. The mechanics of the survey have been described elsewhere \citep{Yang17,Tartaglia2018,Yang19}, along with the recent addition of a second telescope in Australia (for an effective $\sim$12 hour cadence), and improvements to our machine learning search algorithm, and fast telescope triggering infrastructure \citep{Bostroem2020}.

Shortly after discovery, we triggered high cadence observations with the world-wide network of robotic telescopes associated with Las Cumbres Observatory and also the Neil Gehrels $Swift$ Observatory \citep{Gehrels2004}. 
The photometric data from the Las Cumbres Observatory were reduced using the PyRAF-based photometric reduction pipeline {\sc lcogtsnpipe} \citep{Valenti2016}. This pipeline uses a low-order polynomial fit to remove the background and calculates instrumental magnitudes using a standard point-spread function fitting technique. Apparent magnitudes were calibrated using APASS ($B$, $V$, $g$, $r$, $i$) and Landolt ($U$) catalogs. The background contamination was removed by subtracting a reference image and the photometry was extracted from the subtracted images. The $Swift$ UVOT images are reduced using the method described in \cite{Brown2009} using the updated zeropoints of \cite{Breeveld2011}. The multi-band light curves are shown in Figure \ref{fig:photometry} and the magnitudes are listed in Table \ref{tab:photometry}. The $Swift$ photometry is available in the Swift Optical Ultraviolet Supernova Archive (SOUSA; \citealt{Brown2014}).

\begin{figure}
\includegraphics[width=1\linewidth]{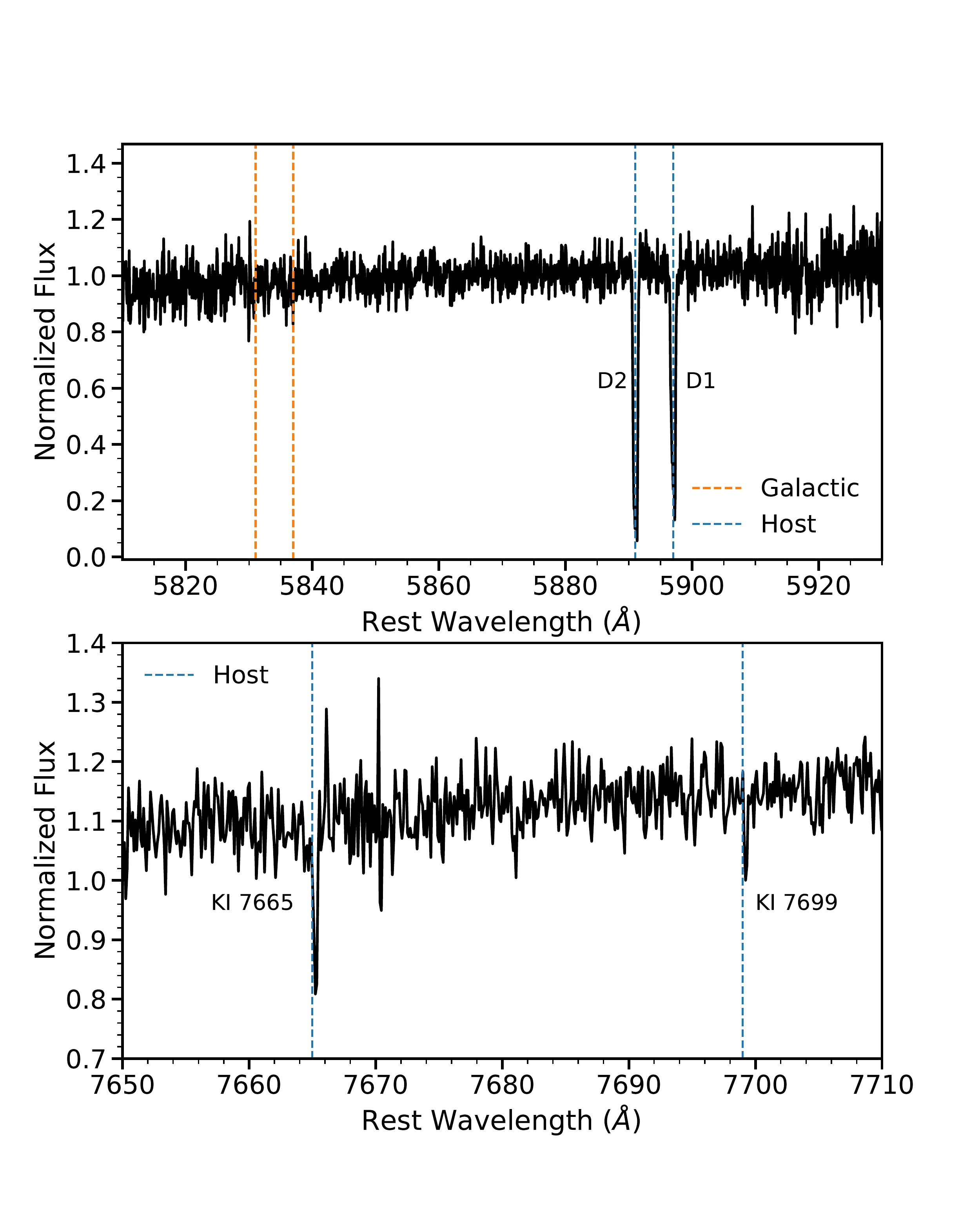}
\caption{An echelle spectrum with a resolution of $\rm R \simeq 40000$ from Magellan/MIKE taken on +18.7 d showing the region around the galactic (dashed orange lines) and host (dashed blue lines) NaID lines (top) and the host KI lines (bottom).
\label{fig:mike_spec}}
\end{figure}

\begin{figure*}
\includegraphics[width=1\linewidth]{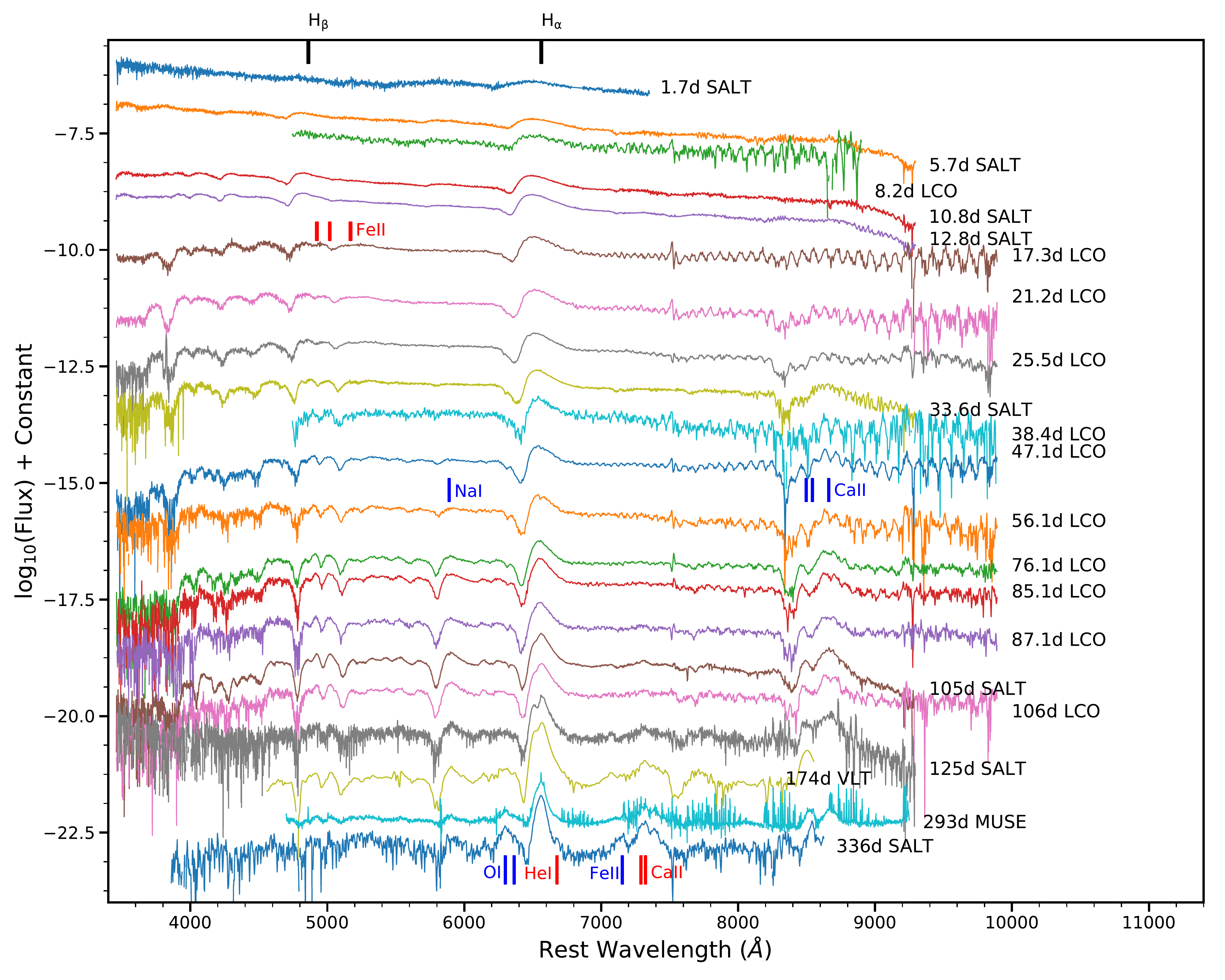}
\caption{The optical spectroscopic evolution of SN~2018cuf from 2 to 336 days after explosion.\label{fig:spec_all}}
\end{figure*}

\begin{figure*}
\includegraphics[width=1\linewidth]{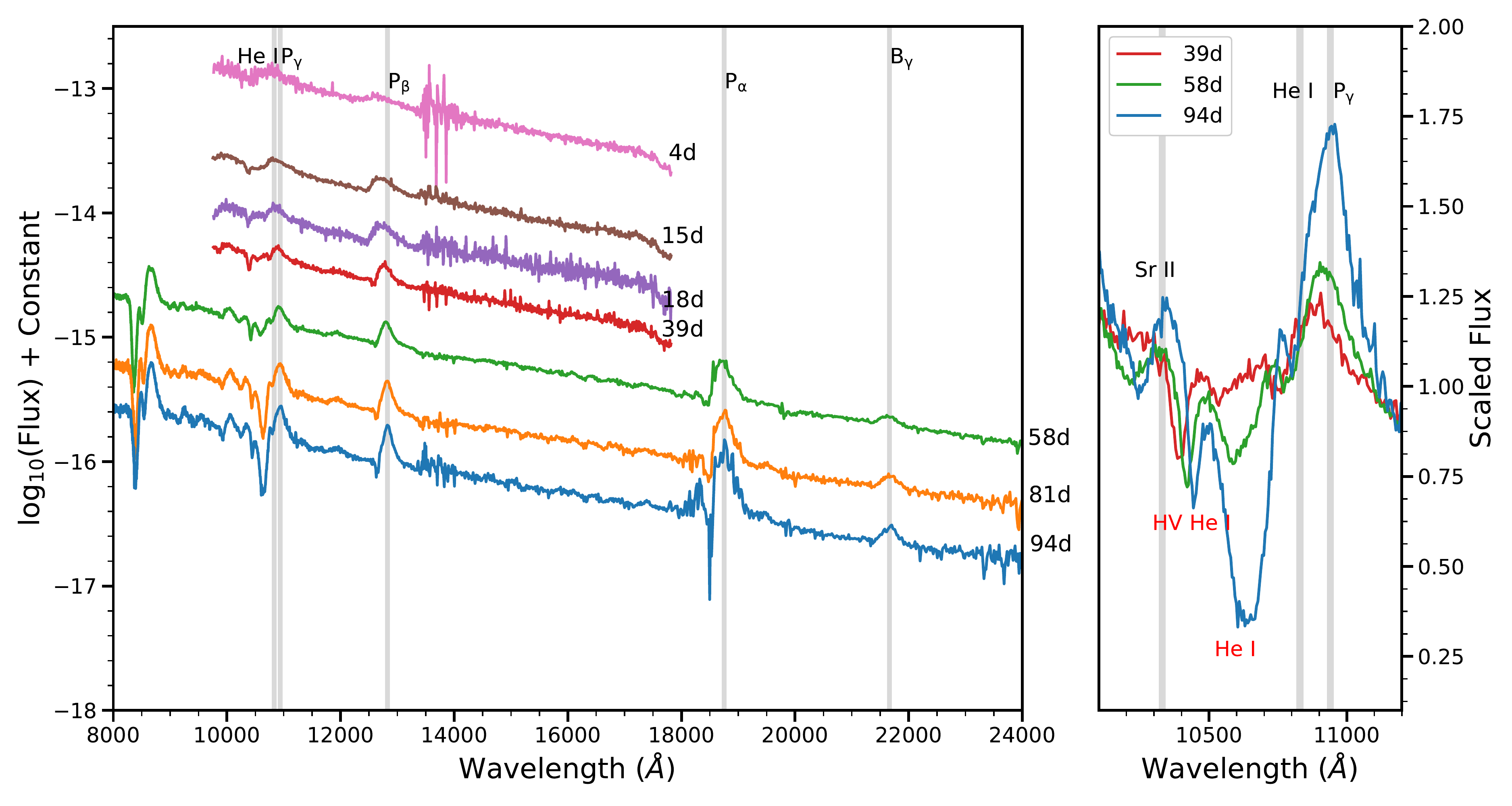}
\caption{Left: Near-infrared spectra of SN~2018cuf from Gemini+FLAMINGOS-2 and Magellan+FIRE. Right: A zoom-in version for spectra at day 39, day 58 and day 94}. The high velocity (HV) He I feature and He I absorption are labelled.\label{fig:NIR_spec}
\end{figure*}

\begin{figure}
\includegraphics[width=1\linewidth]{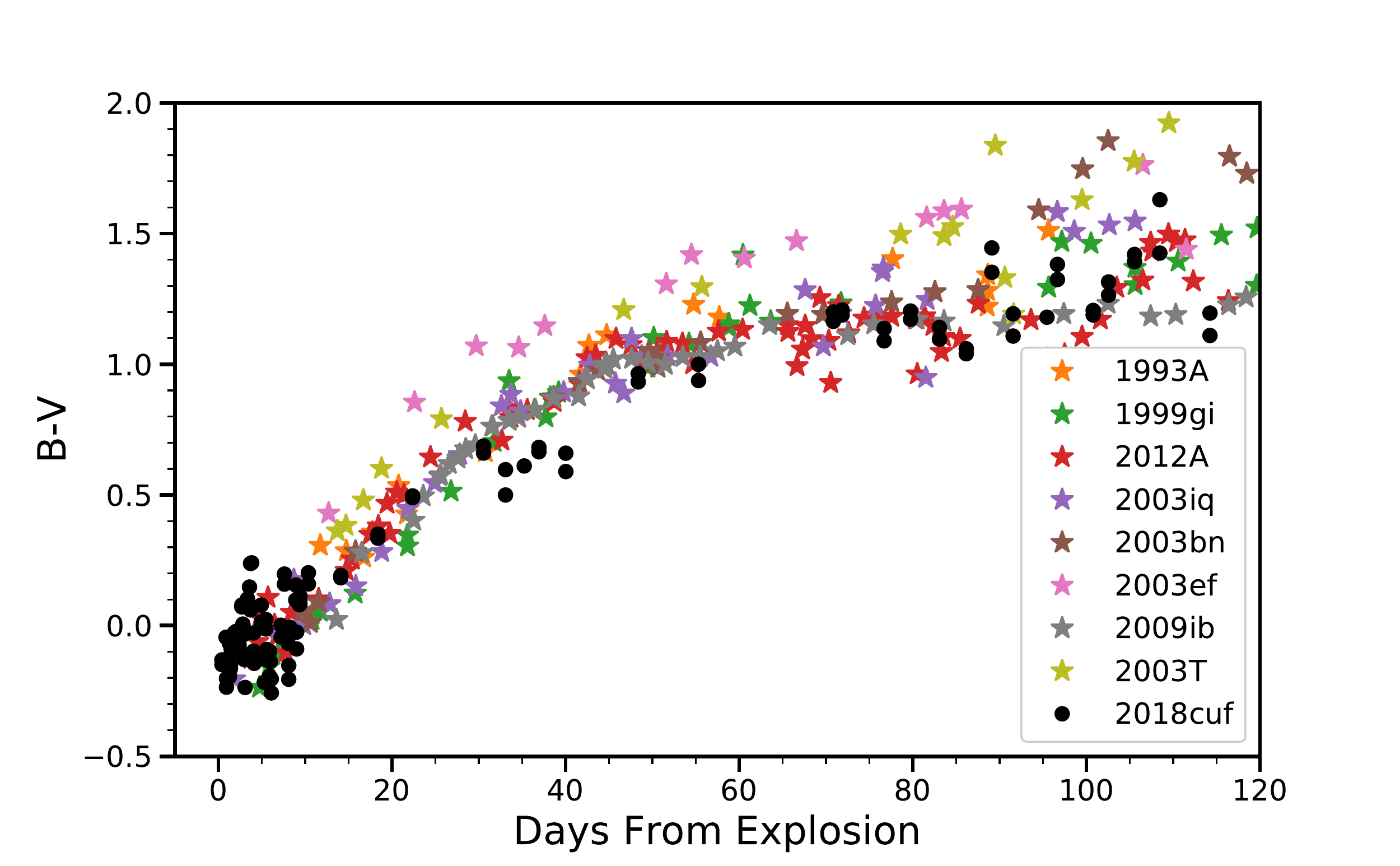}
\caption{Color evolution of SN 2018cuf, after correcting for a total color excess of $E(B-V)_{MW}$ = 0.1373 mag. We also plot a sample of Type II SNe with published reddening estimates (see Section~\ref{sec:reddening} for details), and SN 2018cuf shows a similar color evolution as the other objects. \label{fig:color}}
\end{figure}

The spectroscopic observations of SN~2018cuf started on 2018 June 24 ($\sim$1 day after discovery) and continued through 2019 October 19. A number of optical spectra were collected by the Southern African Large Telescope (SALT), including the first classification spectrum, which classified SN~2018cuf as a young SN Type II \citep{Jha2018}. In addition, many low dispersion optical spectra were obtained by the FLOYDS spectrograph \citep{Brown2013} on the 2m Faulkes Telescope South (FTS) in Australia, and these spectra were reduced following standard procedures using the {\sc FLOYDS} pipeline \citep{Valenti2014}. One optical spectrum was taken with the GMOS instrument \citep{Hook2004,10.1117/12.2233883} at the Gemini South telescope on 2018-06-24 05:33:32 UT, under program GS-2018A-Q-116. GMOS was used in longslit spectroscopy mode with the B600 grating, with a total exposure time of 750 s, and the spectrum was reduced by using the IRAF gemini package. However, this spectrum has a very low S/N, so we did not use it for analysis. We also used FORS2 \citep{Appenzeller1998} at the Very Large Telescope (VLT) with the $\rm GRIS\_150I$ grism and GG435 blocking filter to observe SN 2018cuf on 2018-12-14 00:48:30 UT, as part of the FOSSIL program (Kuncarayakti et al., in prep.). The total exposure time was 2700 s. The data were reduced using EsoReflex software \citep{Freudling2013}. The low dispersion optical spectra are shown in Figure~\ref{fig:spec_all}. There is also one high resolution optical spectrum taken by the Magellan Inamori Kyocera Echelle instrument (MIKE) \citep{Bernstei2003} on the Magellan Clay Telescope (Figure \ref{fig:mike_spec}), and the data was reduced reduced using the latest version of the MIKE pipeline\footnote{\url{https://code.obs.carnegiescience.edu/mike/}} (written by D. Kelson).

Near-infrared (NIR) spectra were taken with the FLAMINGOS-2 instrument (F2, \citealt{Eikenberry2006}) at Gemini South Observatory and the Folded-port InfraRed Echellette instrument (FIRE, \citealt{Simcoe2013}) on the Magellan Baade telescope. The Magellan FIRE spectra were obtained in high throughput prism mode with a 0.6 arcsec slit, giving continuous wavelength coverage from 0.8 to 2.5 $\mu$m.  For the Gemini South F2 spectra, we observed with the JH grism and 0.72 arcsec slit in place, yielding a wavelength range of 1.0--1.8 $\mu$m.  For both the FIRE and F2 data, observations were taken with a standard ABBA pattern for sky subtraction, and an A0V star was observed adjacent to the science exposures for both telluric corrections and flux calibration.  Data for both instruments were reduced in a standard manner as described in \citet{Hsiao19}, and we refer the reader there for the details. The NIR spectra are presented in Figure~\ref{fig:NIR_spec}.
All the spectroscopic observations are listed in Table \ref{tab:spectra} and will be available on WISeREP \citep{Yaron2012}\footnote{\url{http://www.weizmann.ac.il}}.


\section{Reddening and host properties}
\label{sec:reddening}
\subsection{Reddening}

The Milky Way line-of-sight reddening towards SN~2018cuf is
$E(B-V)_{MW}$ = 0.0273 $\pm$ 0.0003 mag \citep{Schlafly2011}.
This low extinction value is also supported by the lack of NaID lines from the Milky Way in the Magellan/MIKE Echelle spectrum taken on 2018 July 12 and shown in Figure \ref{fig:mike_spec}. The equivalent width (EW) of the NaID line is often used to estimate the SN reddening with the assumption that it is a good tracer of gas and dust \citep{Munari1997,Poznanski2012}. 
The measured EW of the host galaxy NaID $\lambda$5890 ($\rm D_2$) and NaID $\lambda$5896 ($\rm D_1$) are 0.677$\rm\AA$ and 0.649$\rm\AA$ respectively. The intensity ratio of $\rm D_2$ to $\rm D_1$ ($\rm D_2$ / $\rm D_1$ $\sim$1) is far from the typical value of 2 we usually observe \citep{Munari1997}, suggesting  that at least $\rm D_2$ may be saturated (see Figure \ref{fig:mike_spec}). Using only $\rm D_1$, we find a  host galaxy extinction of $E(B-V)_{host}$ = 0.699 $\pm$ 0.17 mag.  

\cite{Phillips2013} suggest that the most accurate predictor of extinction is the diffuse interstellar band (DIB) absorption feature at 5780 $\rm\AA$. However, this feature is not clearly present in our high resolution spectrum of SN~2018cuf, suggesting the host galaxy extinction is small, which is inconsistent with the high host reddening derived from NaID lines. \cite{Munari1997} found that [\ion{K}{1}]~$\lambda$7699 can be a better reddening indicator if NaID lines are saturated, so we decide to use this line to estimate the reddening from the host galaxy. The EW of [\ion{K}{1}]~$\lambda$7699 is measured to be 0.03 $\rm \AA$, which corresponds to a host galaxy extinction of $E(B-V)_{host}$ = 0.11 $\pm$ 0.01 mag \citep{Munari1997}. As a sanity check, we also compare the dereddened $B-V$ color evolution of SN~2018cuf to a sample of other similar Type II SNe with published reddening estimates. This includes SN~1993A \citep{Anderson2014,Galbany2016AJ}, SN~1999gi \citep{Leonard2002a}, SN~2003iq \citep{Faran2014}, SN~2003bn \citep{Anderson2014,Galbany2016AJ}, SN~2003ef \citep{Anderson2014,Galbany2016AJ}, SN~2003T \citep{Anderson2014,Galbany2016AJ}, SN~2009ib \citep{Takats2015} and SN~2012A \citep{Tomasella2013}, as is shown in Figure \ref{fig:color}. SN~2018cuf has a similar $V$-band light curve slope after maximum with these selected SNe. \cite{Jaeger2018} found that the color evolution of SNe are related to the slope of the V-band light curve, so these selected SNe should have consistent colors with SN~2018cuf after dereddening. We find that an $E(B-V)_{host}$ $\approx$ 0.11 mag gives us a consistent color evolution with the other objects. Therefore throughout this paper we will adopt an $E(B-V)_{tot}$ = 0.1373 $\pm$ 0.0103 mag, as well as an R$_{V}$ = 3.1 \citep{Cardelli1989}.

The disagreement between host reddening values obtained from the NaID lines versus direct color comparisons to other similar objects is not a unique problem.  \cite{Leonard2002} found a similar situation for SN 1999em, i.e.,  the equivalent width of the sodium lines suggested a high reddening for SN~1999em, but a low value was assumed based on color comparisons. \cite{Phillips2013} also found that NaID gives unreasonably high reddening for some of their objects, while KI line gives a reddening that is consistent with the reddening derived from SN colors.  

\subsection{Host Properties}
Multi Unit Spectroscopic Explorer (MUSE) \citep{Bacon2010} integral field unit (IFU) observations of IC 5092 were taken on 2019 April 12, as a part of the All-weather MUse Supernova Integral-field Nearby Galaxies (AMUSING; \citealt{Galbany2016}) survey. 
MUSE is mounted to the 8.2m Yepun UT4 Very Large Telescope, with a field-of-view of 1\arcmin$\times$1\arcmin and 0.2\arcsec$\times$0.2\arcsec spatial elements, small enough to sample the PSF. See Figure~\ref{fig:sn_image} for an outline of the MUSE footprint.  The spectral coverage is from 4750 to 9300\AA, with a spectral resolution that ranges from R$\simeq$3500 in the blue end, $\simeq$1700 in the red end. 
Four 580s exposures (2320s total exposure time), rotating 90 deg between frames, were taken centered on the South-West side of the galaxy, which covered the SN position and its environment.


We extracted a 3.6$\arcsec$ aperture spectrum centered at the SN position (corresponding to a $\sim$800~pc diameter) to study the properties of the environment. The resulting spectrum is shown in Figure \ref{fig:spec_all}. MUSE observations were performed 293 days after SN~2018cuf's explosion, and some SN features were still visible in the spectrum, with the most pronounced being a broad Balmer H$\alpha$ emission, in addition to a HII region spectrum with narrow emission lines.
To measure the flux of the strongest ionized gas emission lines in that region ([\ion{N}{2}]~$\lambda$6548, H$\alpha$ and [\ion{N}{2}]~$\lambda$6583), we excluded the SN broad component by fitting 4 Gaussians, 3 narrow and 1 broad, simultaneously.
The bluer region of the spectrum was not strongly contaminated by SN features, and we fit single Gaussians to measure the narrow H$\beta$ and [\ion{O}{3}]~$\lambda$5007 emission line fluxes from the ionized gas.

An estimate of the reddening can be obtained from the line-of-sight gas column by the ratio of the Balmer lines, assuming a case B recombination \citep{Osterbrock2006} and a theoretical ratio of H$\alpha$/H$\beta$ = 2.86.
Our lines present a ratio of 4.54, which corresponds to $E(B-V)$ = 0.399$\pm$0.021 mag. This value is not consistent with the reddening estimated from our color comparison (Figure~\ref{fig:color}), and would make the light curves of SN~2018cuf significantly bluer and brighter than other similar Type II SNe. A possible explanation for this disagreement is that the SN is in front of the HII region and not influenced by the dust but the MUSE measurement gets the full column of gas.

With the host galaxy reddening-corrected fluxes we estimate the SN environmental oxygen abundance (O/H) by using the N2 and O3N2 calibrators from \cite{Pettini2004}. We obtain a consistent oxygen abundance of 12+log(O/H) = 8.71 $\pm$ 0.07 dex and 12+log(O/H) = 8.72 $\pm$ 0.08 dex with the N2 and O3N2 calibrators, respectively, both being consistent with solar abundance \citep{Asplund2009}.
We used the H$\alpha$ luminosity to estimate the star formation rate (SFR) at the SN location using the expression provided by \cite{Kennicutt1998}. We obtain a SFR of 0.0014 $\pm$ 0.0001 $\rm
M_{\odot}\ yr^{-1}$, and a SFR intensity of 0.0027 $\pm$ 0.0001 $\rm M_{\odot}\
yr^{-1}\ kpc^{-2}$. To understand where SN~2018cuf stands in the Type II SNe group, we compare the values we derived above with the host properties of all Type II SNe in the PMAS/PPak Integral-field Supernova hosts COmpilation (PISCO) sample \citep{Galbany2018}\footnote{observations are updated to June 2020}. The average host oxygen abundance and SFR intensity for all PISCO Type II hosts are 12+log(O/H) = 8.53 $\pm$ 0.062 dex and 0.013 $\pm$ 0.0014 $\rm M_{\odot}\ yr^{-1}\ kpc^{-2}$, respectively, suggesting that the region around SN~2018cuf has a higher oxygen abundance but a lower SFR intensity than the average of Type II SNe.

\section{Observational Properties} 
\label{sec:obs_properties}

\subsection{Distance}
\begin{figure}
\includegraphics[width=1\linewidth]{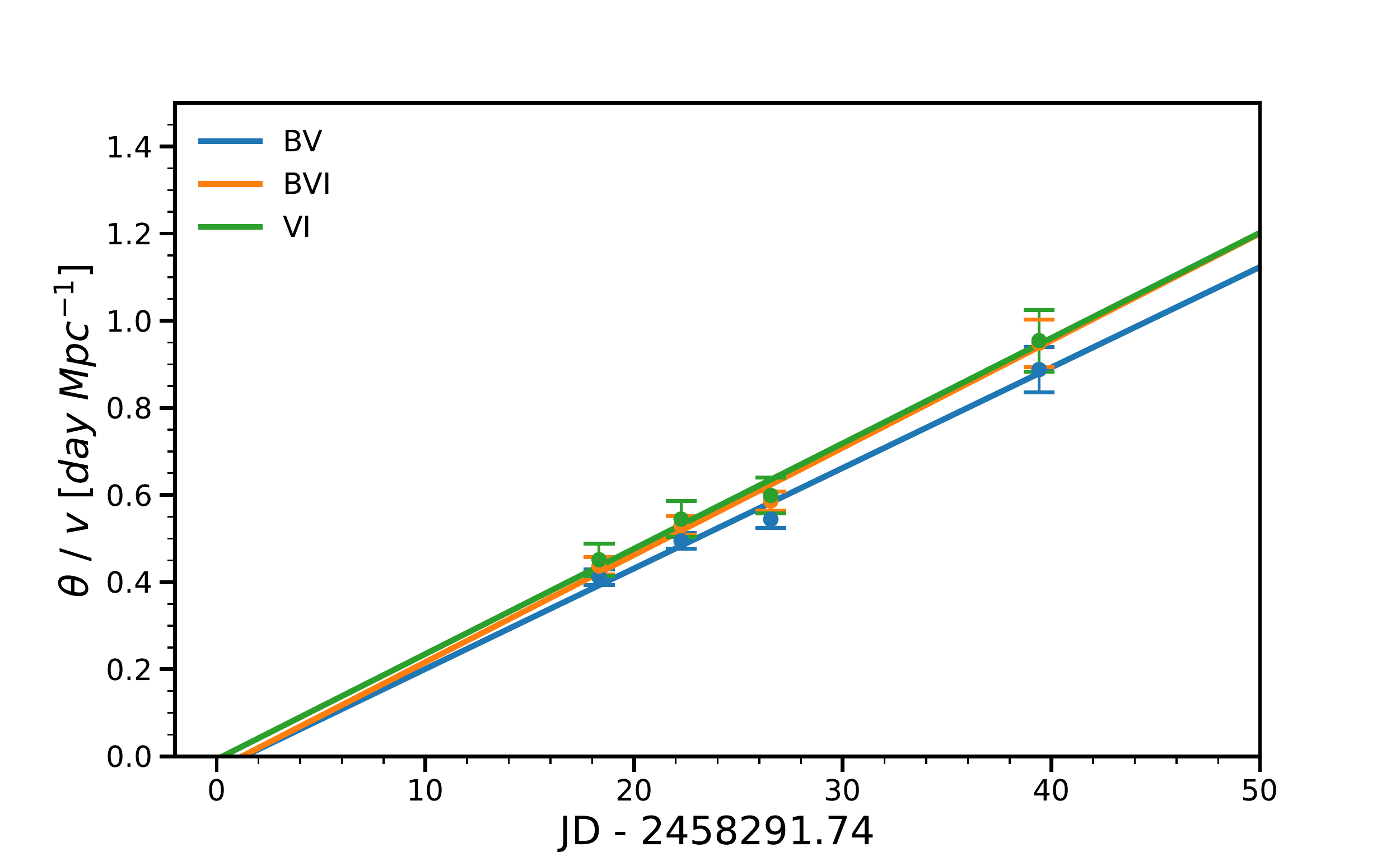}
\caption{EPM fitting for SN 2018cuf using three filter subsets: \{BV\}, \{BVI\} and \{VI\}. The derived distances are 43.3 $\pm$ 5.8 Mpc, 40.6 $\pm$ 5.2 Mpc and 41.3 $\pm$ 7.4 Mpc for the three filter subsets,  respectively, and the weighted average is 41.8 $\pm$ 5.8 Mpc.   \label{fig:epm}}
\end{figure}

The distance to IC 5092 is not well constrained since it has only been measured using the Tully-Fisher relation \citep{Mathewson1992,Willick1997} to be 32.0$\pm$5.8 Mpc.
While the Tully-Fisher relation can be used to measure distances to most spiral galaxies, the intrinsic scatter hinders the accuracy of the measurement for a single galaxy \citep{Czerny2018}. 
One commonly used approach to independently measure the distances to Type II SNe is the expanding photosphere method (EPM), although it requires the object to have well-sampled light curves and spectra. The EPM was first developed by \cite{Kirshner1974} to calculate the distance to Type IIP SNe based on the \cite{Baade1926} method. Assuming that the photosphere is expanding freely and spherically, we can obtain the distance from the linear relation between the angular radius and the expanding velocity of the photosphere using the function:
\begin{equation}
\label{equ: linear}
t = D\,(\frac{\theta}{v_{phot}}) + t_0
\end{equation}
where D is the distance, $t_0$ is the explosion epoch, $\theta$ is the radius of the photosphere (in angular units) and $v_{phot}$ is the velocity of the photosphere. 
Assuming that the photosphere radiates as a dilute blackbody, we combine the multi-band photometry to simultaneously derive the angular size ($\theta$) and color temperature ($T_c$) by minimizing the equation:
\begin{equation}
\epsilon = \sum_{\nu\in S} \{ m_\nu + 5log[\theta\xi(T_c) ] - A_\nu - b_\nu(T_c)\}^2
\end{equation}
where $\xi$ and $b_\nu$ are a dilution factor and synthetic magnitude respectively, and both of them can be treated as a function of $T_c$ \citep{Hamuy2001,Dessart2005}, $A_\nu$ is the reddening, $m_\nu$ is the observed magnitude and $S$ is the filter subsets, i.e., \{BV\}, \{BVI\} and \{VI\}.
We estimate the photospheric velocity by measuring the minimum of the [\ion{Fe}{2}]~$\lambda$5169 P Cygni profile. To accurately estimate the error on this measurement and avoid noise induced local minima, we smooth the spectra with Savitzky-Golay filters \citep{Savitzky1964,Poznanski2010} with different widths, deriving the photospheric velocity for each width. For our distance measurement we use the mean and standard deviation of these velocity measurements. 
After $\sim$ 40 days, the relation between $\theta/ v$ and $t$ is clearly nonlinear \citep{Jones2009} and for this reason we only use four early spectra with clear [\ion{Fe}{2}]~$\lambda$5169 detection, and interpolate the photometry data to the corresponding epoch. The measured velocities are listed in Table \ref{tab:fe_vel}. In order to use the dilution factor derived by \cite{Dessart2005}, we convert $rp$ and $ip$ magnitude to $I$ magnitude by using the equations given by \cite{Lupton2005}. The results for three filter subsets \{BV\}, \{BVI\} and \{VI\} are presented in Figure \ref{fig:epm}. From these measurements, we obtain distances of 43.3 $\pm$ 5.8 Mpc, 40.6 $\pm$ 5.2 Mpc and 41.3 $\pm$ 7.4 Mpc, respectively, and the weighted average is calculated to be 41.8 $\pm$ 5.8 Mpc by using the method described in \cite{Schmelling1995}. In the rest of the paper, we will adopt this value for the analysis. 

\begin{deluxetable}{cc}
\tablenum{1}
\tablecaption{The velocities of [\ion{Fe}{2}]~$\lambda$5169 used in EPM\label{tab:fe_vel}}
\tablewidth{0pt}
\tablehead{
\colhead{Date} & \colhead{[\ion{Fe}{2}]~$\lambda$5169 velocity (km/s)}
}
\startdata
2018-07-10 & $7641.3 \pm 141.8$\\
2018-07-14 & $6664.1 \pm 68.3$\\
2018-07-18 & $6328.8 \pm 79.8$\\
2018-07-31 & $4311.7 \pm 210.1$\\
\enddata{}
\end{deluxetable}

\subsection{Explosion Epoch}
We derive the explosion epoch from the EPM analysis, obtaining similar values from each of the three filter subsets used: JD~2458293.04 $\pm$ 2.88 days in \{BV\}, JD~2458292.95 $\pm$ 2.82 days in \{BVI\}, and JD~2458292.02 $\pm$ 4.22 days in \{VI\}).
The weighted average of these measurements is JD~2458292.81 $\pm$ 3.08, which we adopt as the explosion epoch throughout this paper. 
We note that this is consistent with the tight constraints of the DLT40 survey which place the explosion epoch between JD~2458291.74456 (the last non-detection) and JD~2458292.8609 (the first detection, which is just 0.05 d after the estimated explosion epoch).

As an independent check, we also estimate the explosion epoch by matching the spectra of SN~2018cuf with the spectral templates in SN Identification (SNID) code \citep{Blondin2007}. This method has been used by  \cite{Anderson2014} and \cite{Gutierrez2017a} to constrain the explosion epochs of a sample of Type II SNe.
\cite{Gutierrez2017a} found that with the addition of new spectral templates to the SNID database, the explosion epoch derived from spectral matching may constrain the explosion to within 3.9 days. Following the work of \cite{Gutierrez2017a}, we fixed the fitting range in SNID to 3500-6000 $\rm \AA$ since the blue end of the spectrum contains more information about the SN and evolves more consistently with time for Type II SNe. 
Fixing the explosion epoch to be JD 2458291.91 (from the EPM method), we compare the spectra at 12.10, 19.06, 22.99 and 27.29 days with the SNID templates, where the explosion epochs are given by the EPM. The top five matches are then averaged to compute the epoch of the spectra and the error is given by the standard deviation. The epochs of spectra derived from this method are 10.84 $\pm$ 1.87, 17.82 $\pm$ 4.74, 25.02 $\pm$ 4.65 and 31.74 $\pm$ 6.79 days, respectively, consistent with the spectral epochs inferred from the EPM.

\section{Photometric evolution}
\label{sec:photo_evol}
The full multi-band light curves of SN~2018cuf are shown in Figure \ref{fig:photometry}.
The $V$-band light curve shows an initial rise to reach a maximum brightness of $M_V$~=~$-16.73 \pm 0.32$ mag on JD~2458300.537, $\sim$9 days after the date of explosion. A plateau of approximately constant brightness follows due to the hydrogen envelope recombination that extends up to roughly day 112. The other filters show similar trends with bluer bands peaking slightly earlier, and redder bands later. Following the plateau phase, the light curves show an unusually slow drop and finally settles onto a linear decline phase.

\begin{figure}
\includegraphics[width=1\linewidth]{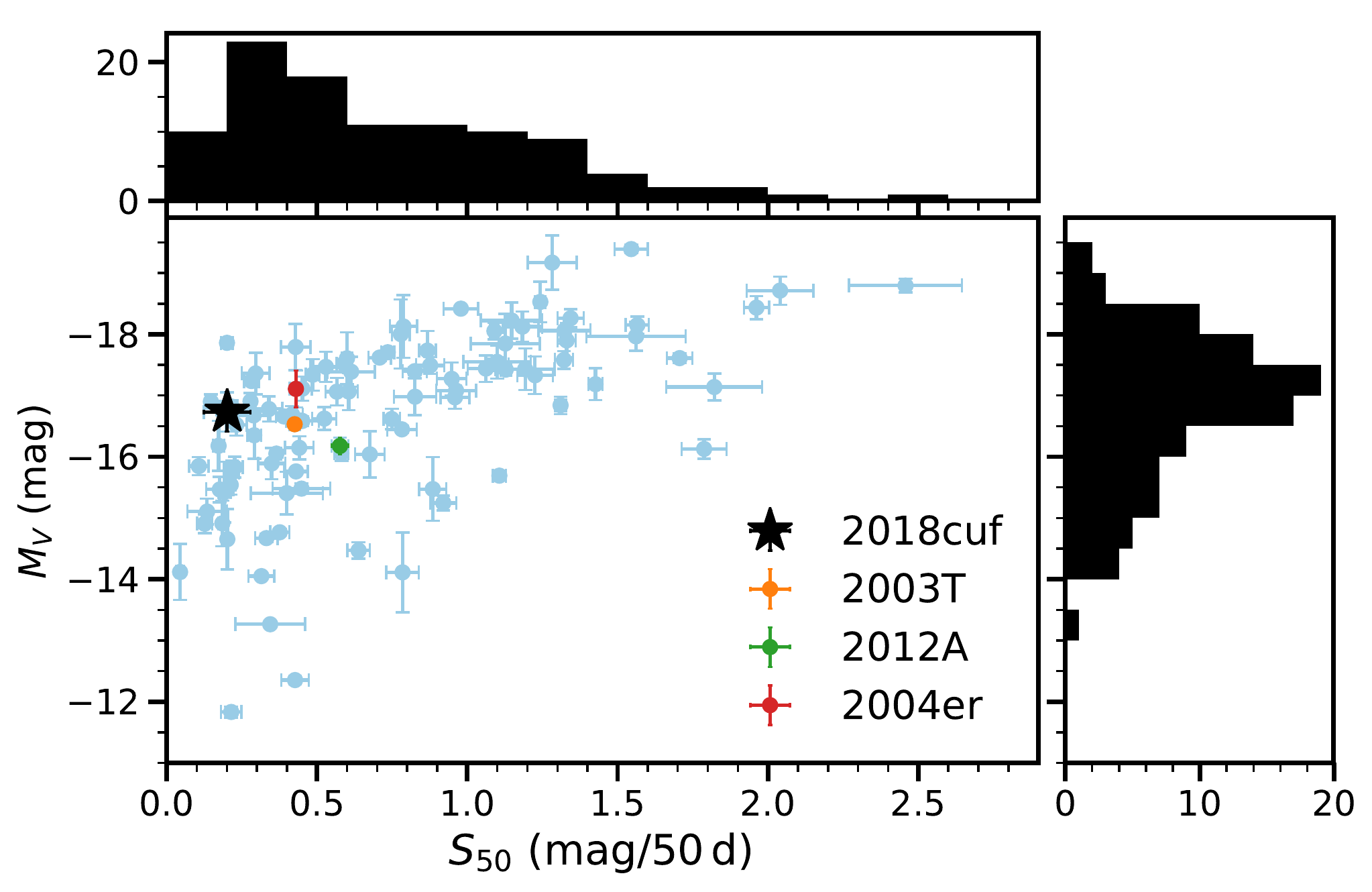}
\caption{The $M_{V}$ compared to $S_{50V}$ for SN~2018cuf and a sample of Type II SNe. The SNe in this sample are from \cite{Anderson2014} and \cite{Valenti2016}, and are available in SNDAVIS. In this plot, IIP-like SNe are usually towards the left, and IIL-like SNe are towards the right. SN~2018cuf, SN~2003T, SN~2004er, and SN 2012A are highlighted with different colors, and the detailed discussion for there four objects can be found in Section \ref{sec:photo_evol}.\label{fig:S50_MV}}
\end{figure}

After maximum brightness, Type II SNe light curves exhibit a wide range of properties. In order to understand where SN~2018cuf lies in the family of Type II SNe, we measure several light curve parameters and compare them with other Type II SNe.
One of the most studied parameters is the rate of decline after maximum light, which is used to classify sub-types of Type II SNe into SNe IIP and SNe IIL (``L'' for linear).
Statistical analyses of Type II SNe also point out that there is a correlation between the decline rate and the maximum absolute magnitude \citep{Li2011,Anderson2014,Valenti2016,Galbany2016AJ}.
Following \cite{Valenti2016} we measure the slope of the light curve per 50 d in $V$ band (S$_{50V}$) of SN 2018cuf. We find S$_{50V}$=0.21 $\pm$ 0.05 $\rm mag\ (50d)^{-1}$, which combined with $M_{V}$ (see Figure \ref{fig:S50_MV}) places SN~2018cuf nicely within the region of Type IIP SNe.

\begin{figure}
\includegraphics[width=1\linewidth]{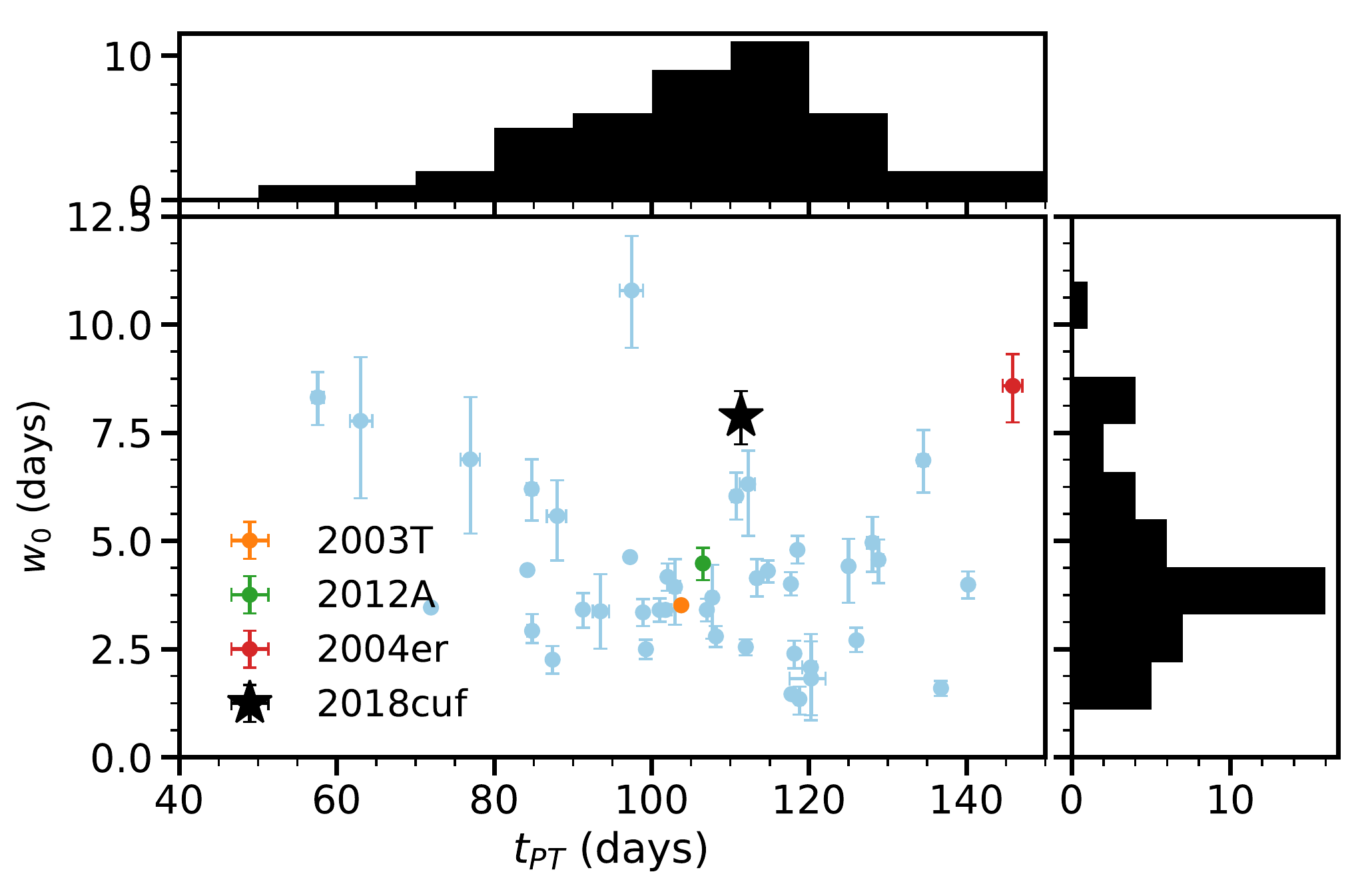}
\caption{Comparison of $t_{\rm PT}$ and $w0$ for $V$ band as described in the text. 
A large $\omega0$ implies a shallow post-plateau slope. SN~2018cuf, SN~2003T, SN~2004er, and SN~2012A are highlighted with different colors.  \label{fig:t_pt_w0}}
\end{figure}

\begin{figure}
\includegraphics[width=1\linewidth]{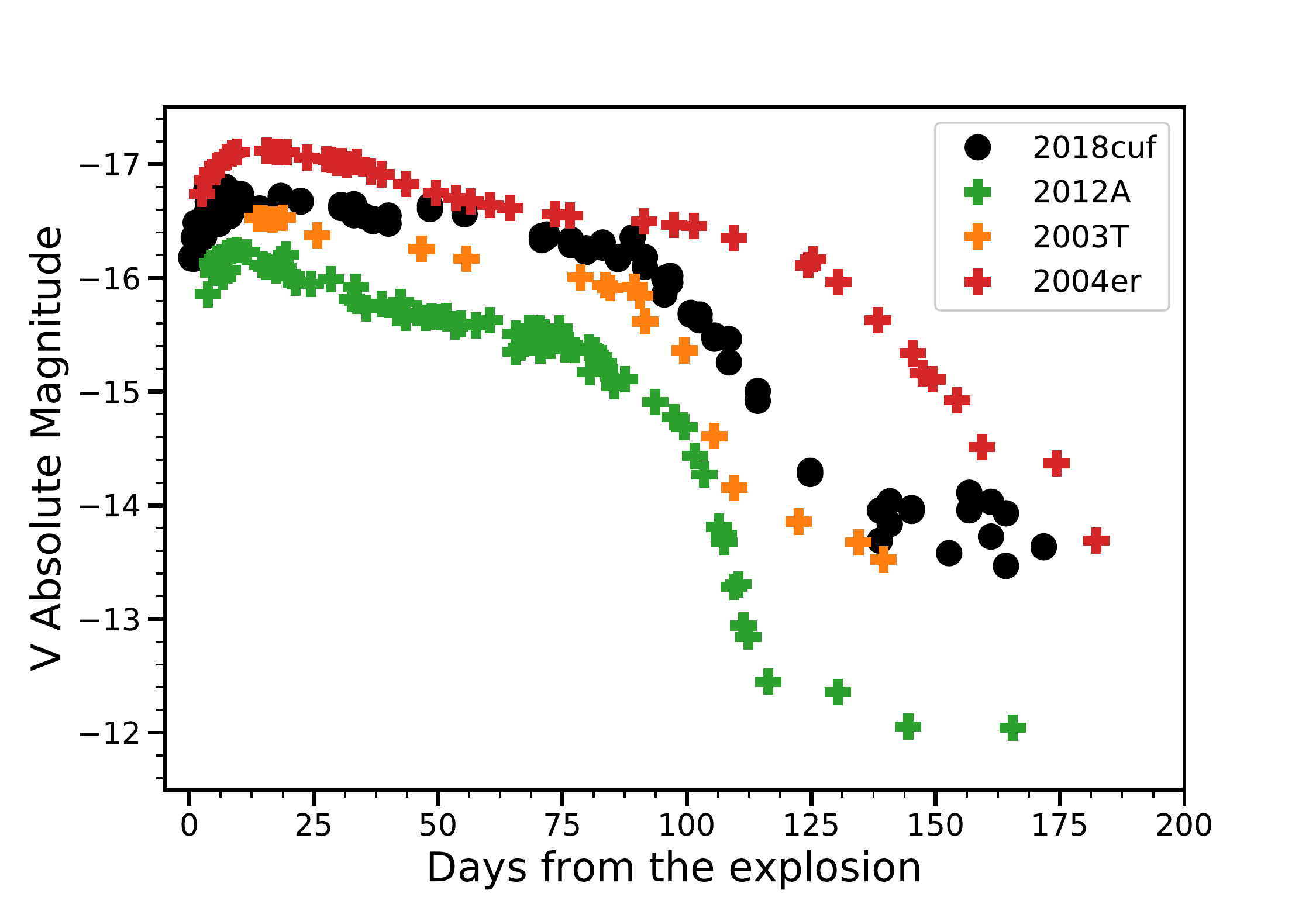}
\caption{$V$-band light curves of SN~2018cuf, SN~2003T, SN~2012A, and 
SN~2004er compared in absolute magnitude. As can be seen in Figure \ref{fig:S50_MV}, SN~2012A and SN~2003T peak at roughly the same magnitude as SN~2018cuf but have a slightly steeper post-plateau slope. SN~2004er has a shallow post-plateau slope, similar to SN~2018cuf, but with a much brighter absolute magnitude and a much longer plateau phase. \label{fig:photo_V}.}
\end{figure}

After $\sim$100 days, the light curves of Type II SNe transition from being powered primarily by the recombination in the photosphere, to being powered by the radioactive decay of $\rm{{}^{56}Ni\rightarrow{}^{56}Co\rightarrow{}^{56}Fe}$. 
This period, known as the fall from plateau, can be characterized as a Fermi-Dirac function \citep{Olivares2010,Valenti2016}: 
\begin{equation}\label{eqn:lc_pars}
    y(t) = \frac{-a0}{1+e^{(t-t_{\rm PT})/ \omega0}} + (p0 \times (t-t_{\rm PT})) + m0,
\end{equation}
where $t_{\rm PT}$ refers to the length of the plateau, $\omega0$ indicates the slope of the light curve during the post-plateau phase (a large $\omega0$ implies a small slope), and $a0$ is the depth of the drop. 
We fit the $V$-band light curve using the package emcee \citep{Foreman2013} and the best-fitting values are found to be $t_{\rm PT}$ = $112.24^{+0.71}_{-0.68}$ d, $\omega0$ = $7.87^{+0.64}_{-0.59}$ d and $a0$ = $1.99^{+0.052}_{-0.049}$ mag.
We find that SN 2018cuf has one of the highest $\omega0$ values in our sample of Type II SNe from the SNDAVIS database\footnote{\url{http://dark.physics.ucdavis.edu/sndavis/transient}} (see Figure \ref{fig:t_pt_w0}), indicating that the slope of the fall from plateau is shallower than most type II SNe. Another SN with a  slow fall from plateau is SN~2004er \citep{Anderson2014} (see Figure \ref{fig:photo_V}), but sparse data on the tail and a lack of multi-color observations make further comparisons difficult.


The effect of $\rm^{56}Ni$ mixing on the SN light curve, in particular in its relation to the fall from plateau, has been studied by many authors \cite[e.g.][]{Kasen2009,Bersten2011,Goldberg2019}.
It is possible that the slow fall from plateau is 
related to a low mixing of the $\rm^{56}Ni$ distribution in the ejecta at the moment of explosion. 
For instance, \cite{Goldberg2019} produced several model light curves with different $\rm^{56}Ni$ distributions (see their Figure 10),
showing that insufficient mixing of $\rm^{56}Ni$ 
results in a shallow slope in the post-plateau phase. 
Alternatively, increasing the total mass of $\rm^{56}Ni$ can also lead to a shallower fall from plateau (e.g., see Figure 2 in \citealt{Kasen2009} and Figure 13 in \citealt{Goldberg2019}).  

Which of these two effects, $\rm ^{56}Ni$ mixing or total $\rm^{56}Ni$ mass, is more important to explain the shallow slope of SN~2018cuf is unclear. To try to disentangle these effects, we identify two other type IIP SNe, SN~2012A and SN~2003T, in the literature that have either a similar progenitor (the progenitor of SN~2018cuf is discussed in Section \ref{sec:progenitor}) or similar light-curve parameters to SN~2018cuf. The $V$-band light curve of SN~2018cuf is compared with those of other SNe in Figure \ref{fig:photo_V}. All three objects are spectroscopically similar with roughly the same maximum absolute magnitude (see Figure \ref{fig:S50_MV}) and plateau length. 
The progenitor of SN~2012A has been well studied \citep{Tomasella2013, Utrobin2015, Morozova2018}, and it has similar progenitor mass, radius and explosion energy to the progenitor of SN~2018cuf but a lower nickel mass. Additionally, SN~2012A has a similar maximum magnitude to SN~2018cuf but a steeper fall from plateau (see Figure \ref{fig:photo_V}).  
By comparing SN~2018cuf with SN~2012A, we may then conclude that the shallow slope of the fall from plateau of SN~2018cuf
is due to the larger nickel produced by SN~2018cuf.
On the other hand, a different conclusion is supported by comparing SN~2018cuf with SN~2003T.
The nickel mass of SN~2003T is very similar to SN~2018cuf according to its tail magnitude, while the fall from plateau of SN~2003T is much faster than that of SN~2018cuf, suggesting that a low degree of nickel mixing in 
SN~2018cuf could also contribute to the shallow fall from plateau.
In addition, the $\rm^{56}Ni$ mass of SN~2018cuf is measured to be 0.04 (0.01) $\rm M_{\odot}$ (see Section \ref{sec:progenitor}). This is consistent with the amount of nickel typically produced in SNe II \citep{Anderson2014,Muller2017,Anderson2019}.
Since the $\rm^{56}Ni$ mass of SN~2018cuf is typical for the Type II family, it is more likely that the shallow slope is due to mixing; however, both scenarios are possible and we are unable to conclusively disentangle the effects.

\begin{figure}
\includegraphics[width=1\linewidth]{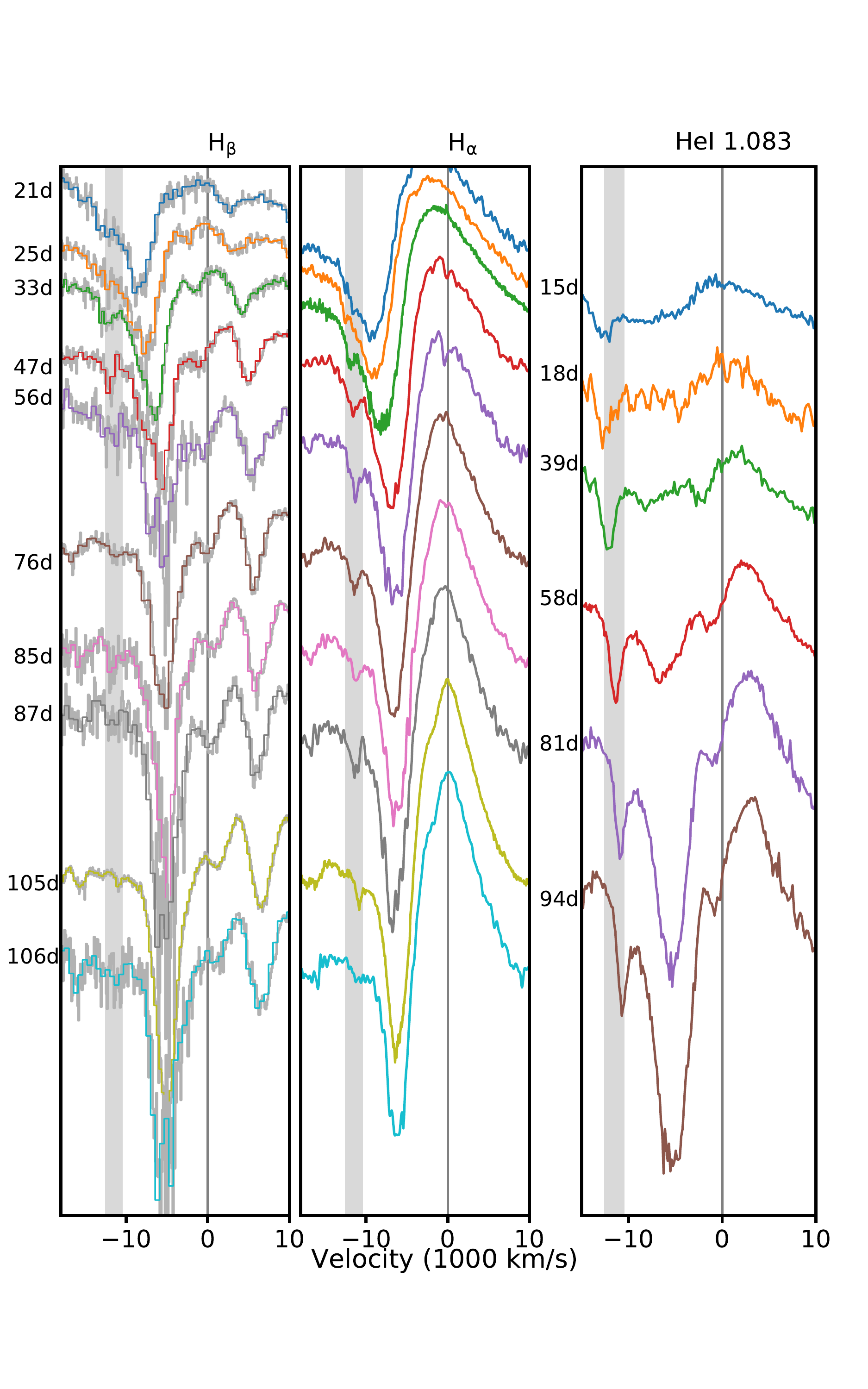}
\caption{The evolution of Cachito features in $\rm H{\alpha}$, $\rm H{\beta}$ and He I $\lambda1.083\ \rm \mu m$ during the photospheric phase. In the left panel, the spectra from FLOYDS and SALT are binned to 9 $\rm \AA\ pixel^{-1}$ and 5 $\rm \AA\ pixel^{-1}$, respectively, and the grey background lines are the original spectra. The shaded area marks -12500km/s to -10500km/s. The Cachito features in all three lines show consistent velocities, supporting their presence as HV features.
\label{fig:HV_spec}}
\end{figure}

\section{Spectroscopic Evolution} 
\label{sec:spec_evol}
\subsection{Optical spectra}
The optical spectroscopic evolution of SN~2018cuf is shown in Figure \ref{fig:spec_all}. The early spectrum shows a blue continuum with a broad $\rm H \alpha$ line clearly detected. Over time, the spectra become redder and develop hydrogen Balmer lines with P~Cygni features. The [\ion{Fe}{2}]~$\lambda\lambda\lambda$4924, 5018 and 5169 lines, good tracers for the photospheric velocity, can be seen after day 17.
Other typical features such as [\ion{Ca}{2}]~$\lambda\lambda$3934, 3968, the \ion{Ca}{2} infrared triplet~$\lambda\lambda\lambda$ 8498, 8542, 8662, and NaID~$\lambda\lambda$5890, 5896 also appear in emission as the SN evolves. During the nebular phase, strong [\ion{Ca}{2}]~$\lambda\lambda$7291, 7324 emission lines emerge along with 
[\ion{Fe}{2}]~$\lambda$7155,  [\ion{He}{1}]~$\lambda$6678 and [\ion{O}{1}]~$\lambda\lambda$ 6300, 6364.

Interestingly, from day 105 to day 174, a small notch appears on the $\rm H{\alpha}$ profile with a velocity of $\sim$ 1000km/s, and its origin is unclear.
One possibility is that this feature is from dust formation either in the ejecta or in CSM interaction. The signatures of dust formation have been detected in many Type IIn SNe. Type IIn SN2010jl shows notches or double peaked profiles at an earlier stage and later shows more dominant blue-wings \citep[see Extended Data Figure 3 of][]{gall14}. Type IIn SN1998S also show a notch feature in their broad emission lines \citep{mauerhan12}. However, for SN~2018cuf, this notch feature emerges starting at day 106, and the temperature of the ejecta may still be too hot for dust formation. On the other hand, the feature is not detected in spectra after day 293, which is hard to reconcile with dust formation. By comparing H$\alpha$ with other hydrogen line profiles, we do not find any evidence of red-side attenuation for lines that occur at bluer wavelengths, as is expected for dust creation. For these reasons, we can not unambiguously attribute this feature to dust formation and equally rule out the possibility of dust formation.

The evolution of $\rm H \alpha$ and $\rm H \beta$ lines during the photospheric phase is shown in Figure \ref{fig:HV_spec}. Starting at day 22, an extra absorption line can be seen on the blue side of the $\rm H \alpha$ and $\rm H \beta$ P Cygni absorption lines, becoming more obvious by day 34. These lines have been studied in many SNe II and have been most often associated with  [\ion{Si}{2}]~$\lambda$6355 when seen at early phases ($<$ 30 d) and to HV hydrogen if seen at a later phase (50-100 d) (e.g., \citealt{Chugai2007}; SN~2005cs, \citealt{Pastorello2006}; SN~2009bw, \citealt{Inserra2012}; SN~2013ej, \citealt{Valenti2014}).  This ambiguous absorption feature is often referred to as the  ``Cachito'' feature \citep{Gutierrez2017a}.
In the case of SN~2018cuf, because this feature appears at roughly 30 days it is likely associated with HV hydrogen, an interpretation that is confirmed by the additional presence of the HV feature in $\rm H \beta$ at similar velocities.

\subsection{NIR Spectra}
The NIR spectra from day 4 to day 94 are plotted in Figure \ref{fig:NIR_spec} and show an evolution typical of Type II SN. The first spectrum at day 4 is nearly featureless with weak Paschen lines but by day 15 these features have strengthened and [\ion{He}{1}]~$\lambda$10830 has also appeared. Both Pa~$\alpha$ and Br~$\gamma$ lines can be seen in our spectra after day 58.

In general, the line evolution in NIR spectra is consistent across Type II SNe. However, \cite{Davis2019} points out that Type II SNe can be classified as spectroscopically strong or weak based on the pseudo-equivalent width (pEW) of the [\ion{He}{1}]~$\lambda$10830 absorption line and the features seen in the spectra.
They find that SNe with weak \ion{He}{1} (pEW $<$ 50\AA) are slow-declining Type IIP SNe and that SNe with strong He I (pEW $>$ 50\AA) correspond to the faster-declining Type IIL SNe class. 
Interestingly, SN~2018cuf seems to be an exception to this rule. The pEWs of \ion{He}{1} absorption for days 39, 58, 81, and 94 are 4.9 $\rm\AA$, 25.2 $\rm\rm\AA$, 82.5 $\rm\AA$ and 105 $\rm\AA$, respectively, which makes it hard to classify it as either a strong or weak SN based on the pEW alone. In addition to smaller pEW, weak SNe usually show the $\rm P{\gamma}/Sr\ II$ absorption feature at earlier epochs ($\sim$ 20 days after explosion), and are accompanied by a HV \ion{He}{1} feature. For SN~2018cuf, the $\rm P{\gamma}/Sr\ II$ absorption feature shows up at day 18, consistent with a weak SN. Additionally, there is clearly an extra absorption feature on the blue side of \ion{He}{1} (see the right panel of Figure \ref{fig:NIR_spec} or Figure \ref{fig:HV_spec}). Other than HV He I, this feature could also be explained as [\ion{C}{1}]~$\lambda10693$. However, the lack of other \ion{C}{1} lines in the NIR spectra makes it unlikely that this feature is originating from [\ion{C}{1}]~$\lambda10693$. We also note that the velocity of HV He I matches the velocity of HV H$\alpha$ and HV H$\beta$ in optical spectra, which further strengthens our conclusion that this feature can be interpreted as HV He I. Although the pEW of SN~2018cuf is higher than the 50 $\rm \AA$ limit used in \cite{Davis2019}, the presence of $\rm P{\gamma}/Sr\ II$ at early phase and HV He I suggests that our object still falls into the weak SN II category. This implies that the 50 $\rm \AA$ limit from \cite{Davis2019} is probably too low.

\cite{Chugai2007} proposed that HV absorption features, like those seen in SN~2018cuf, come from the interaction between the circumstellar (CS) wind and the SN ejecta. They argue that there are two physical origins of HV absorption: enhanced excitation of the outer layers of unshocked ejecta, which contributes to the shallow HV absorption in the blue side of H$\alpha$ and [\ion{He}{1}]~$\lambda$10830, and the cold dense shell (CDS), which is responsible for the HV notch in the blue wing of H$\rm \alpha$ and H$\rm \beta$. For the former case, the $\rm H\beta$ HV is not expected to be seen due to the low optical depth in $\rm H\beta$ line-forming region, whereas in the latter case the HV Cachito can form in both $\rm H \alpha$ and $\rm H\beta$. For SN~2018cuf, the presence of Cachito features in $\rm H \alpha, H \beta$, and [\ion{He}{1}]~$\lambda$10830 supports the CDS interpretation, but does not completely rule out the first scenario.

$\rm Pa\beta$ and $\rm Pa\gamma$ were also investigated to look for the presence of HV features. However, the existence of other strong lines around $\rm Pa\gamma$ makes it difficult to identify a HV feature if present, and there is no HV feature in the blue side of $\rm Pa\beta$. \cite{Chevalier1994} suggested that the temperature of the CDS should be low enough that this region is dominated by low-ionization lines, which causes Pa$\beta$ absorption to form in a low optical depth region and may explain the absence of HV feature in Pa$\beta$.

\section{Progenitor Properties}
\label{sec:progenitor}

\subsection{Nickel Mass}
The nebular phase of Type II SNe is driven by radioactive decay $\rm^{56}Ni \rightarrow$ $\rm^{56}Co \rightarrow$ $\rm^{56}Fe$. If the $\gamma$-rays produced by this process are completely trapped by the ejecta, the bolometric luminosity at late times can be used to estimate the amount of $\rm ^{56} Ni$. Since our photometry after $\sim$100 days does not cover the full SED, we use two different methods to derive the $\rm ^{56}Ni$ mass.
The first method is to calculate the pseudo-bolometric luminosity of SN~2018cuf and compare it with the pseudo-bolometric light curve of SN~1987A.
Assuming SN~2018cuf and SN~1987A have the same normalized SED, the nickel mass is given by \cite{Spiro2014}: 
\begin{equation}
\label{eq:ni}
M_{\rm Ni} = 0.075 \times \frac{L_{SN}}{L_{87A}} \rm M_{\odot}
\end{equation}
where $L_{SN}$ and $L_{87A}$ are the pseudo-bolometric luminosity of SN~2018cuf and SN~1987A, respectively. For the pseudo-bolometric luminosity, we follow the method described by \cite{Valenti2008}. The observed magnitudes are converted to flux at each band and integrated by using Simpson's rule, which uses a quadratic polynomial to approximate the integral. Photometric data from day 135 to day 170 are used to calculate the pseudo-bolometric luminosities of SN~2018cuf and SN~1987A by using passbands \{BVgri\} and \{BVRI\}, respectively, resulting in a $\rm ^{56}Ni$ mass of $0.037^{+0.003}_{-0.002}\rm\ M_{\odot}$.

An alternative approach to estimate the nickel mass is to compute a full-band bolometric light curve by performing a black body fit to all available filters at each photometric epoch and integrating the black body. The advantage of this method is that it does not require the assumption that SN~2018cuf and SN~1987A have the same normalized SED, although the approximation to a black body may not be completely valid due to the line blanketing in the UV bands. The $\rm ^{56}Ni$ mass derived from this approach is $0.042^{+0.045}_{-0.008}\rm\ M_{\odot}$.
Given the limitations of each method, we choose to use the pseudo-bolometric luminosity method to estimate the $\rm ^{56}Ni$ mass, but take the difference between the results from the two methods as an indicator of the uncertainty of the measurement. The final nickel mass is conservatively estimated to be $M_{\rm Ni} = 0.04\ (0.01)~\rm M_{\odot}$. By comparing the pseudo-bolometric light curve of SN~2018cuf with that of SN~1987A, we found that the decline rate of SN~2018cuf in the radioactive tail is either consistent with or slightly faster than $\rm ^{56}Co$ decay. It is hard to be sure which one is the case here due to the lack of data in the radioactive tail. If the decline rate of SN~2018cuf is slightly faster than $\rm ^{56}Co$ decay, the $\rm ^{56}Ni$ mass we derived here could be treated as a lower limit.

\subsection{Progenitor Mass}
Progenitor mass is a fundamental parameter of a SN, and can be constrained by using multiple techniques. In this section, we derive the progenitor mass of SN~2018cuf from nebular spectra and hydrodynamic light curve modelling.

\subsubsection{From Nebular Spectroscopy}
During the nebular phase, spectra can provide useful information about the inner structure of a SN. At this stage, the ejecta has become optically thin, revealing the core nucleosynthesis products. The strength of the [\ion{O}{1}]~$\lambda\lambda6300, 6364$ doublet in the nebular spectra has been found to be a good indicator of progenitor mass \citep{Jerkstrand2014}. By comparing the intensities of [\ion{O}{1}]~$\lambda\lambda6300, 6364$ with theoretical models during this phase, the progenitor mass can be well constrained. \cite{Jerkstrand2014} modelled the nebular spectra for 12, 15, 19, and 25 $\rm M_{\odot}$ progenitors at different phases. They start with the SN ejecta evolved and exploded with KEPLER \citep{Woosley2007} and use the spectral synthesis code described in \cite{Jerkstrand2011} to generate the model spectra. 

\begin{figure}
\includegraphics[width=1\linewidth]{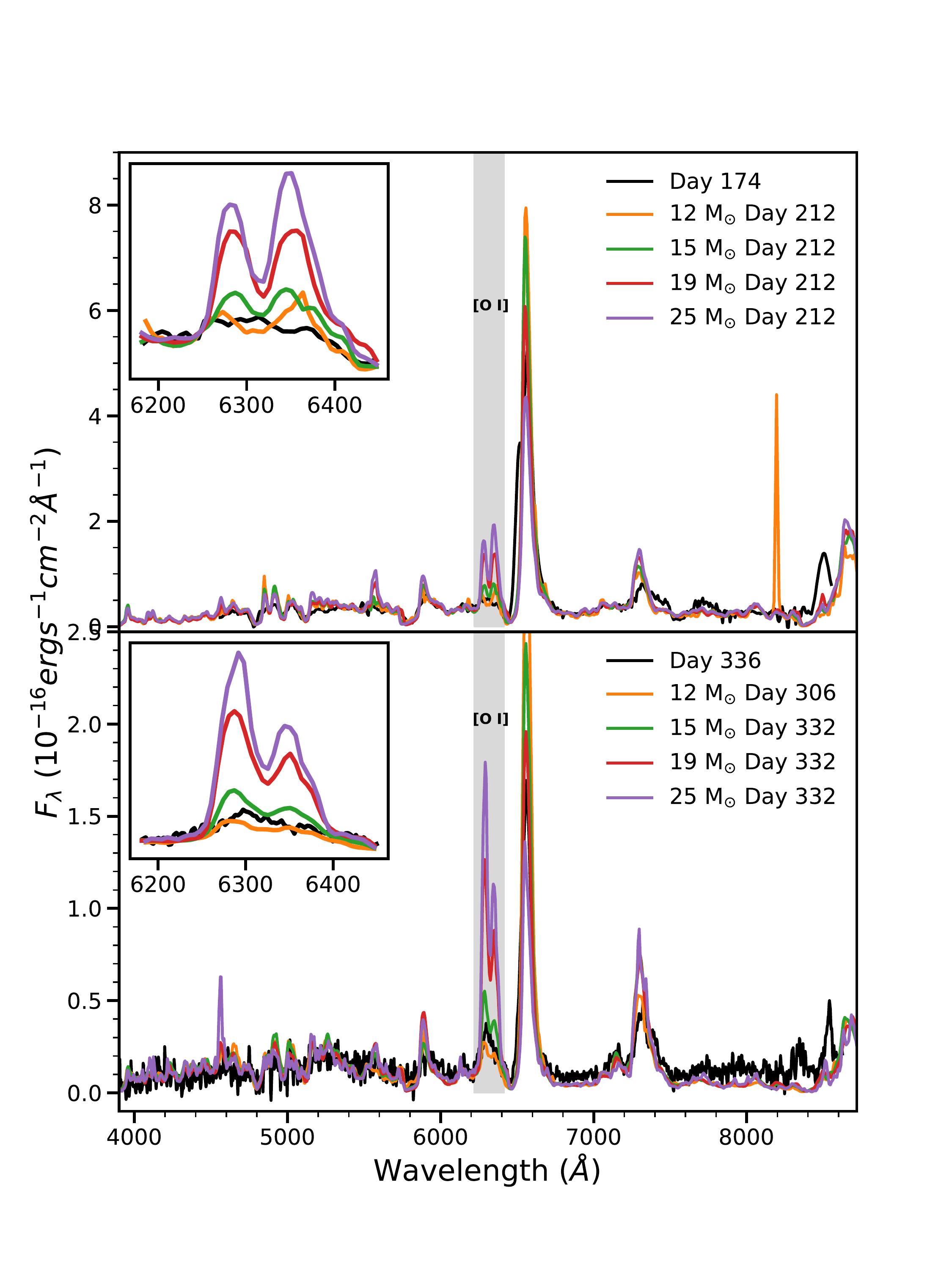}
\caption{Comparison of the nebular spectra of SN~2018cuf from day 174 and day 336 with four models at similar epochs. The insets show the [\ion{O}{1}] doublet, which is a good indicator of progenitor mass.  From this line, we estimate the progenitor mass to be between 12 and 15 $\rm M_{\odot}$ as discussed in the text. \label{fig:nebular}}
\end{figure}

Although we have six nebular spectra for SN~2018cuf taken from day 174 to day 483, four of them are contaminated by the host galaxy. Therefore, we only compare the nebular spectra of SN~2018cuf at day 174 and day 336 with the models computed by \cite{Jerkstrand2014} in Figure \ref{fig:nebular}. We scale the nebular spectra taken at day 174 and day 336 to $r$-band photometry, and the models have been scaled to the observed spectrum so that they have the same integrated flux. We find that the strength of \ion{O}{1} in our spectrum is between the 12 $\rm M_{\odot}$ and the 15 $\rm M_{\odot}$ models, which implies the progenitor mass of SN~2018cuf is likely in this range. 

Synthetic nebular spectra can also be used to give an independent estimate of the nickel mass \citep{Jerkstrand2018, Bostroem2019}. By using the scale factors we used to scale the model spectra, the nickel mass can be derived using the following relation from \cite{Bostroem2019}:
\begin{equation}
    \frac{F_{obs}}{F_{mod}} = \frac{d^2_{mod}}{d^2_{obs}}\frac{M_{^{56}Ni_{ obs}}}{M_{^{56}Ni_{mod}}}exp\left(\frac{t_{mod}-t_{obs}}{111.4}\right)
\end{equation}
where $F_{obs}$ is the total observed flux and $F_{mod}$ is the total flux from the model spectrum. $d_{obs}$ is the distance of the SN and $d_{mod}$ = 5.5 Mpc is the distance used to compute the model; $M_{^{56}Ni}$ indicates nickel mass for the SN ($M_{^{56}Ni_{obs}}$) and the model ($M_{^{56}Ni_{mod}} = 0.062 \rm M_{\odot}$), and $t_{obs}$ and $t_{mod}$ is the phase of the spectra for the observation and model, respectively. 
We then derive nickel masses of $0.040^{+0.004}_{-0.003}$ $\rm M_{\odot}$ and $0.060^{+0.006}_{-0.016}$ $\rm M_{\odot}$ for day 174 and day 336, respectively.
These values are consistent with what we get in the previous subsection, where we measure the nickel mass from the radioactive decay tail photometry when the SN just falls from the plateau (day 135 -- 170).

\begin{figure}
\includegraphics[width=1\linewidth]{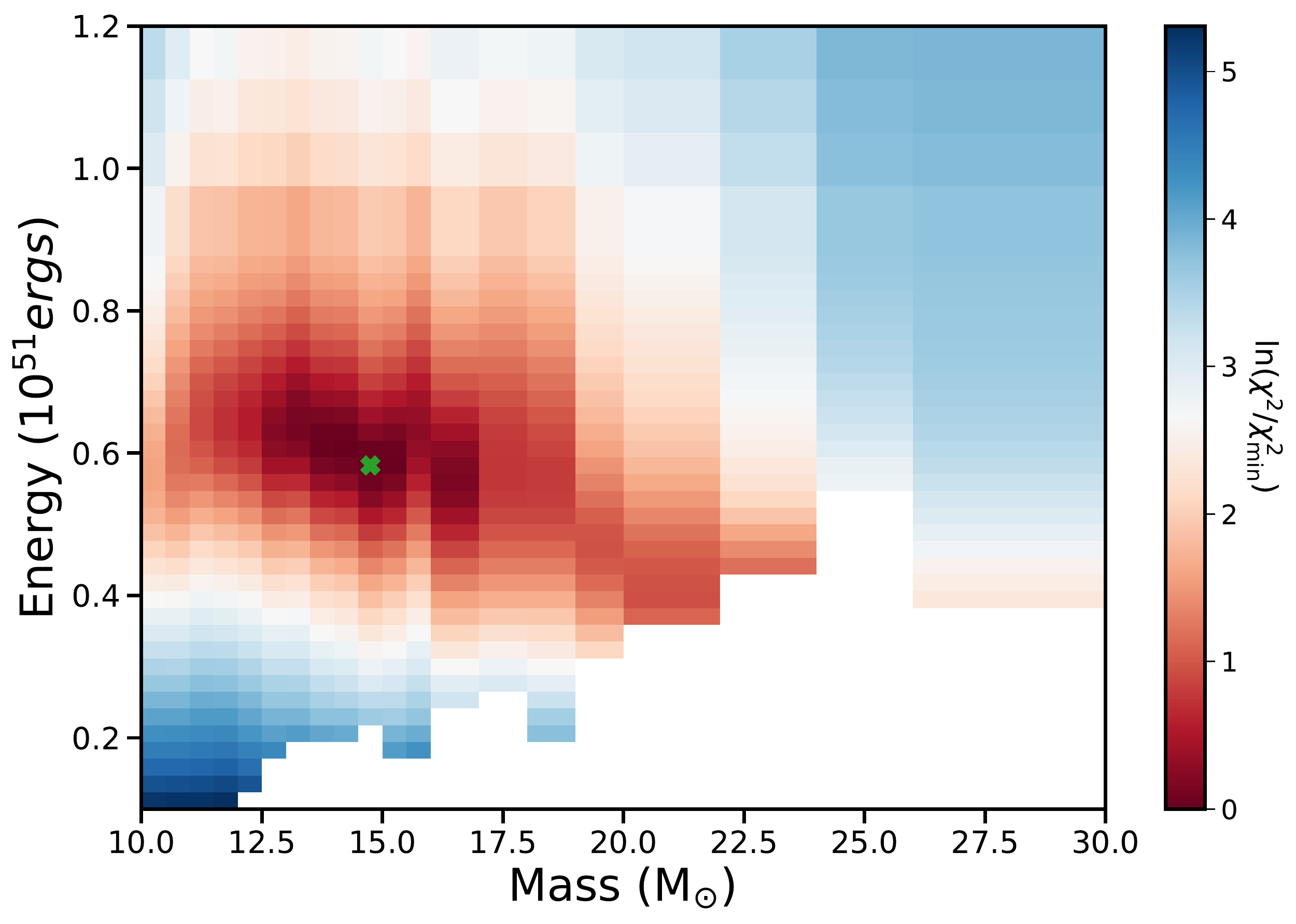}
\includegraphics[width=1\linewidth]{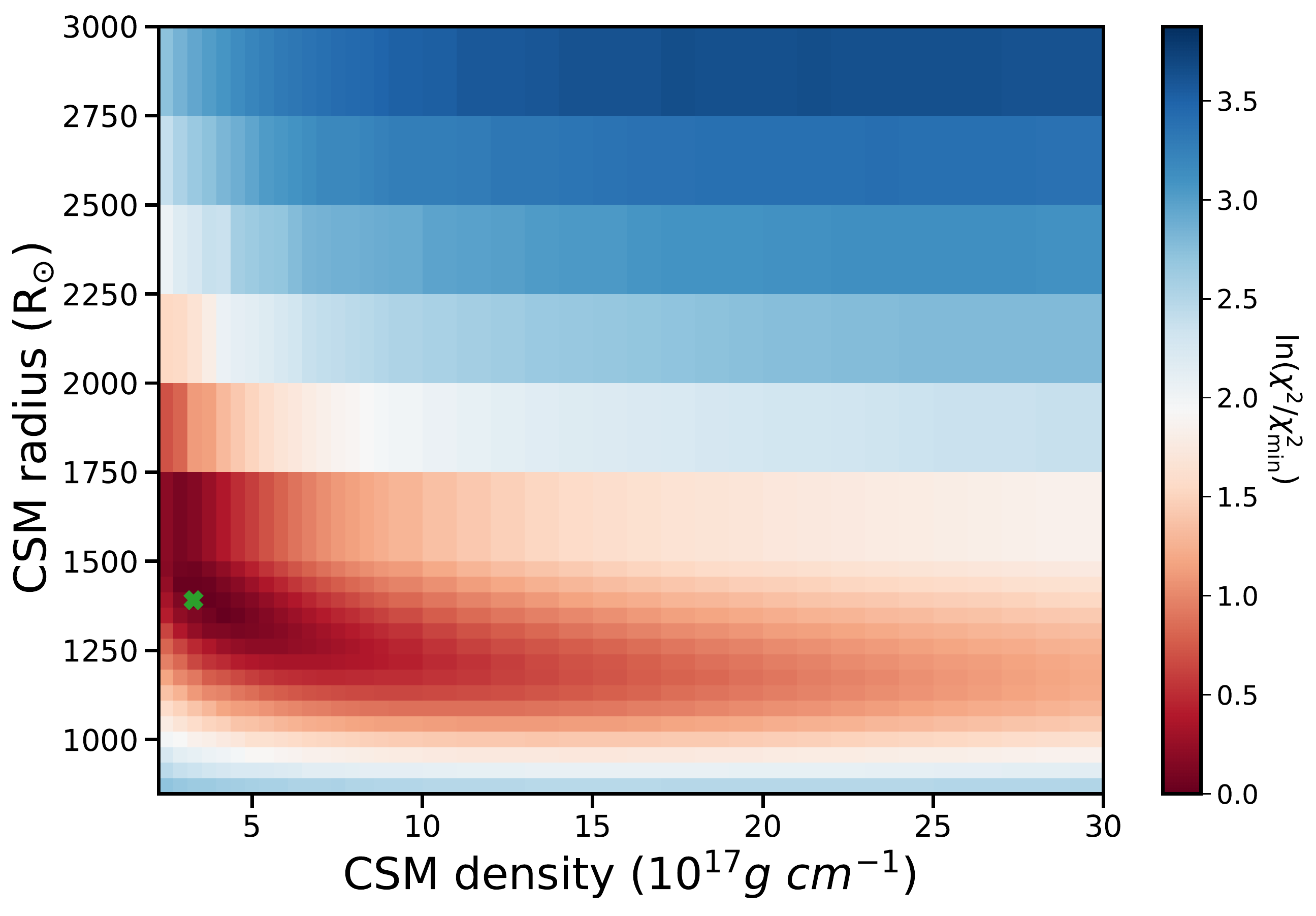}
\includegraphics[width=1\linewidth]{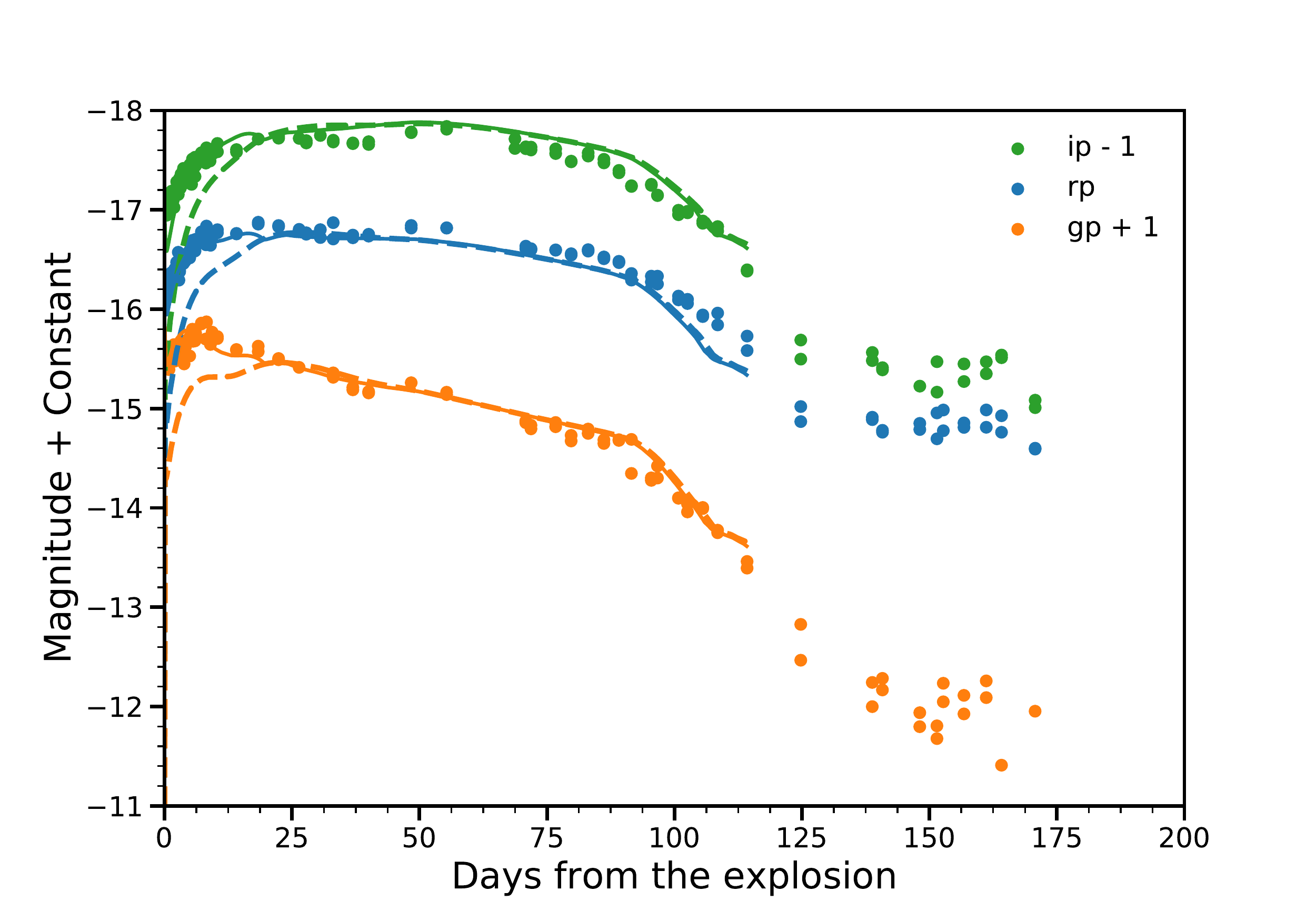}
\caption{Top and middle: The color indicates the $\chi^{2}$ value, and green cross represents the best-fitting SNEC model. Bottom: Dots are the observational data, while different colors represent different bands. Solid lines and dashed lines are the best-fitting SNEC model with and without CSM respectively. \label{fig:snec}}
\end{figure}

\subsubsection{From Hydrodynamic Modelling}
An alternative way of constraining the mass of the progenitor is to compare the light curves to hydrodynamic models \citep[e.g.][]{Utrobin2015,Utrobin2017,Morozova2017,Morozova2018,Paxton2018,Goldberg2019,Martinez2019}. We have used the SN Explosion Code (SNEC; \citealp{Morozova2015}), 
an open-source hydrodynamic code for core-collapse SNe, to constrain the progenitor parameters of SN~2018cuf.
SNEC assumes diffusive radiation transport and local thermodynamic equilibrium (LTE), which are valid assumptions from shock breakout through the end of the plateau. However, as the SN becomes nebular 
this assumption breaks down. 
For this reason, we compare our light curve only out to $t_{\rm PT}=112.24$ days with the SNEC models.
Our inputs of evolved progenitor stars for the SNEC code are  the non-rotating solar metallicity RSG models generated from the {\sc kepler} code and described in \cite{Sukhbold2016}. A steady-state wind with a density profile
\begin{equation}
    \rho(r) = \frac{\dot M}{4\pi r^2 \nu_{wind}}=\frac{K}{r^2}
    \label{eqn:wind}
\end{equation}
is also added above these models to explore the effect of CSM on light curves, where $\dot M$ is the wind mass-loss rate and $\nu_{wind}$ is the wind velocity. We will use parameter $K$ to describe the constant wind density, which extends up to radius $R_{ext}$.
For each explosion, SNEC takes a variety of progenitor and explosion parameters as input and then generates a bolometric light curve and, assuming blackbody radiation, the optical light curves.
We followed the approach of \cite{Morozova2017}, exploring variations in progenitor mass ($M$), nickel mass, explosion energy ($E$), $K$ and $R_{ext}$, fixing the Nickel mass to $M_{\rm Ni} = 0.04\ \rm M_{\odot}$, which we obtained from the tail photometry. We note that the degree of $\rm^{56}Ni$ mixing can also be a free parameter in SNEC models. However, the SNEC model can not reproduce the light curves well during the fall from plateau since the radiation diffusion approach used in SNEC is no longer valid during and after this period, so we are not able use SNEC to explore the effect of $\rm^{56}Ni$ mixing on the post-plateau light curves. \cite{Morozova2015} also found that the light curves generated by SNEC are not sensitive to the degree of $\rm^{56}Ni$ mixing, so we fixed the initial $\rm^{56}Ni$ mixing, and mixed $\rm^{56}Ni$ up to 5 $\rm M_{\odot}$ in the mass coordinates. 

\cite{Morozova2018} points out that only the early phase of the light curve is dominated by CSM, so it is possible to adopt a two-step approach to fit the light curves. In the first step, we evaluate the fit for the part of the light curve that is mostly dominated by the hydrogen-rich envelope and vary only $M$ and $E$. The fitting range is chosen to be between the end of $s1$ (37.19 days from explosion for SN~2018cuf) and $t_{\rm PT}$, where $s1$ is the initial steeper slope of the light curve. This allows us to determine the best fit progenitor mass and explosion energy. In the next step, we fix the progenitor mass and explosion energy found in step one and we explore the influence of CSM, varying $K$ and $R_{ext}$ and fitting the whole light curve through $t_{\rm PT}$. This substantially reduces the number of models needed to explore the parameter space, allowing us to search over a finer grid in each parameter. 

At each stage, the best fit model was determined by interpolating the models to the observed epochs in $g, r, i$ filters and minimizing the $\chi^2$ over these filters. For the first stage, the range of parameters considered is $10\ \rm M_{\odot} <$ M $< 30\ \rm M_{\odot}$ and $0.1 < E < 1.2$ (in unit of $10^{51}\ $ ergs). We obtain the best fit of $M = 14.5\ \rm M_{\odot}$, which corresponds to a 827 $\rm R_{\odot}$ progenitor star from Table 2 of \cite{Sukhbold2016}, and $E = 5.71\times10^{50}\ $ ergs as shown in the upper panel of Figure \ref{fig:snec}. 
In the next step, the CSM parameter range is set to be $2 < K < 20$ (in unit of $10^{17}\ g\ cm^{-1}$) and $827\ \rm R_{\odot}<$ $R_{ext} <$ $3000\ \rm R_{\odot}$, and the fitting range also includes the early part of the light curve, i.e., we fit the light curves from the explosion to $t_{\rm PT}$. The result is presented in the middle panel of Figure \ref{fig:snec}, and the best-fitting model is $K$ = $3.1\times10^{17}\ g\ cm^{-1}$ and $R_{ext} = 1369\ \rm R_{\odot}$.  In the bottom panel of Figure \ref{fig:snec}, we show the light curves of the best-fitting models with and without dense CSM. The progenitor mass (14.5 $\rm M_{\odot}$) we get from SNEC model is in good agreement with what we get from the synthetic nebular spectra analysis (12 - 15 $\rm M_{\odot}$), and this is a moderate mass for a Type II SN. It should be noted that we did not fit the model photospheric velocity with the ejecta velocity derived from the [\ion{Fe}{2}]~$\lambda$5169 line. Both \cite{Goldberg2019,Goldberg2020} pointed out that fitting the ejecta velocities inferred from the [\ion{Fe}{2}]~$\lambda$5169 line can barely break the degeneracy between the explosion properties, so we choose not to fit the ejecta velocity in our SNEC modelling.

Previous workers \citep{Morozova2017,Morozova2018,Bostroem2019} found that there is a strong degeneracy between the density profile and the external radius of CSM, and the total mass of CSM derived from the fits is more robust. If we adopt the progenitor radius of 827 $\rm R_{\odot}$ as the inner CSM radius, the total CSM mass of our best fit model is found to be 0.07 $\rm M_{\odot}$ by integrating Equation \ref{eqn:wind} over $r$. If we interpret this wind as that of a typical RSG, adopting a wind speed of 10 $\rm km~s^{-1}$, the mass-loss rate would be 0.06 $\rm M_{\odot}\ yr^{-1}$ within a timescale of 14 months, much higher than the steady winds observed in RSGs \citep{Smith2014}. The possible explanation is that such dense CSM may originate from pre-SN outbursts due to the late-stage nuclear burning in the stellar interior \citep{Quataert2012,Smith2014a,Fuller2017,Ouchi2019,Morozova2020}. Due to the presence of dense CSM around the SN, it is also expected to see flash signatures in the early spectrum \citep{Yaron2017,Nakaoka2018,Rui2019}. However, such a signature is not found for SN~2018cuf, which may imply that the dense CSM is very close to the progenitor, consistent with the small CSM radial extent derived from the SNEC model.

\subsection{Shock Cooling Model}
After the shock breakout, the SN emission is dominated by shock cooling, and carries useful information about the radius and pre-explosion evolution of the progenitor star system. \cite{Sapir2017} updated the model presented by \cite{Rabinak2011} and found that the photospheric temperature and bolometric luminosity during the early phase for a SN with a RSG progenitor (convective envelope with n = 3/2) can be written as:
\begin{eqnarray}
T(t) = T_1 * t_d^{-0.45}\\
L(t) = L_1\ exp\left[-\left(\frac{1.67 * t_d}{t_{tr}}\right)^{0.8}\right] * t_d^{-0.17},
\end{eqnarray}
where $T_1$ and $L_1$ are the temperature and the luminosity $\sim 1$ day after the explosion respectively, $t_d$ is the time from explosion, $t_{tr}$ is the time when the envelope is starting to become transparent. 
We apply this model to SN~2018cuf, which was discovered well before $t_{tr}$, using the code developed by \cite{Hosseinzadeh2018,griffin_hosseinzadeh2020}. This Markov Chain Monte Carlo (MCMC) routine is adopted to give the posterior probability distributions of $T_1$, $L_1$, $t_{tr}$ and $t_0$ simultaneously, where $t_0$ is the explosion epoch. This analytical model is only valid for T $<$ 0.7 eV, and we have checked that the final fitting results satisfy this condition. The MCMC converge to an explosion epoch MJD~58287.8 $\pm$ 0.2 (or JD~2458288.3 $\pm$ 0.2), which is about three day earlier than our last non-detection (JD~2458291.74). An explosion epoch earlier than our first non-detection is possible as the SN may be below our detection limits shortly after explosion. However, in order to fit the $U$- and $V$-band light curves, we require $\rm t_{tr}\ =\ 10000~days$, which is unphysically late. For this reason we do not attempt to derive progenitor or explosion parameters using this method. The inability of this method to fit the blue part of the light curve has been noted by several authors \citep{Arcavi17,Hosseinzadeh2018}. One possible reason for the fitting failure could be that there is a CSM-ejecta interaction around the progenitor, which is supported by the light curve modelling as we discussed in the last subsection. In addition, the effect of UV-band line blanketing is underestimated in the model spectrum, so that assuming black body radiation can not well reproduce the light curves in UV bands.

\section{conclusions} 
\label{sec:conclusion}
In this paper, we have presented the spectroscopic and photometric observations of SN~2018cuf in the galaxy IC~5092. The object was discovered by the DLT40 survey within $\sim$1 day of explosion, and the well-sampled light curves and spectra from GSP were used to constrain the progenitor properties.
In general, SN~2018cuf is consistent with other Type II SNe, while it has a relatively slow fall from the plateau, which could be a result of insufficient mixing of $\rm ^{56}Ni$ or high $\rm ^{56}Ni$ mass. During the plateau phase, we identified HV features in $\rm H{\alpha}$, $\rm H{\beta}$ and $\rm He\ I\ \lambda10830$, suggesting interaction between ejecta and CSM.

We use the EPM method to derive a distance of 41.8 $\pm$ 5.8 Mpc to SN~2018cuf and an explosion epoch of JD 2458292.81 $\pm$ 3.08, which is confirmed by SNID and consistent with the last nondetection from DLT40. From the pseudo-bolometric luminosity of the radioactive decay tail, the nickel mass is found to be 0.04 (0.01) $\rm M_{\odot}$, which is further confirmed by the nickel mass derived from nebular spectra. SNEC modelling is used to determine the progenitor parameters finding a progenitor mass of $ 14.5\ \rm M_{\odot}$ with an explosion energy of $E \approx  5.71\times10^{50}\ \rm erg$, and a CSM mass of $M_{CSM}\approx$ 0.07 $\rm M_{\odot}$. The progenitor mass from SNEC is in good agreement with what we get from nebular spectral modelling ($ 12-15\ \rm M_{\odot}$). The dense CSM inferred from SNEC modelling may imply that the progenitor experienced some outbursts due to the late-stage nuclear burning before explosion. We also tried to apply the shock cooling model to the early light curve, but find it yields unphysical results. From the SNEC model, we infer significant CSM around SN~2018cuf, which could be a main reason for the fitting failure, since the shock cooling model is no longer valid in the presence of dense CSM. In addition, the underestimate of the effect of UV-band line blanketing for the model spectra may also contribute to the failure of model fitting.

We found that, at least for this single object, hydrodynamical modelling and nebular spectral modelling give consistent progenitor mass. In the future, with more and more young SNe detected, we will be able to investigate the systematic bias for these techniques and finally have the ability to better understand the progenitors of Type IIP SNe.

\acknowledgments
We would like to thank Daniel Kasen and Nir Sapir for beneficial discussions. 
Research by Y.D., and S.V., and K.A.B is supported by NSF grants AST-1813176. Research by DJS is supported by NSF grants AST-1821967, 1821987, 1813708, 1813466, 1908972, and by the Heising-Simons Foundation under grant \#2020-1864.
This work makes use of observations from the Las Cumbres Observatory network. DAH, JB, and DH are supported by NSF grant AST-1911225 and NASA Swift grant 80NSSC19k1639.
L.G. was funded by the European Union's Horizon 2020 research and innovation programme under the Marie Sk\l{}odowska-Curie grant agreement No. 839090. This work has been partially supported by the Spanish grant PGC2018-095317-B-C21 within the European Funds for Regional Development (FEDER).
The SALT observations presented here were made through Rutgers University programs 2018-1-MLT-006 and 2019-1-MLT-004 (PI: Jha); supernova research at Rutgers is supported by NSF award AST-1615455.
The GMOS observation, GS-2018A-Q-116, was conducted via the time exchange program between Gemini and the Subaru Telescope.
E.Y.H. and S.D. acknowledge the support provided by the National Science Foundation under Grant No. AST-1613472.
IA is a CIFAR Azrieli Global Scholar in the Gravity and the Extreme Universe Program and acknowledges support from that program, from the European Research Council (ERC) under the European Union’s Horizon 2020 research and innovation program (grant agreement number 852097), from the Israel Science Foundation (grant numbers 2108/18 and 2752/19), from the United States - Israel Binational Science Foundation (BSF), and from the Israeli Council for Higher Education Alon Fellowship.
H.K. was funded by the Academy of Finland projects 324504 and 328898.
Based on observations collected at the European Southern Observatory under ESO programme 0102.D-0356.
X.W. is supported by National Natural Science Foundation of China (NSFC grants 11633002 and 11761141001), and the National Program on Key Research and Development Project (grant no. 2016YFA0400803). 
L.W. is sponsored (in part) by the Chinese Academy of Sciences (CAS), through a grant to the CAS South America Center for Astronomy (CASSACA) in Santiago, Chile.

This research has made use of the NASA/IPAC Extragalactic Database (NED), which is operated by the Jet Propulsion Laboratory, California Institute of Technology, under contract with the National Aeronautics and Space Administration.

\vspace{5mm}
\facilities{CTIO:PROMPT, Las Cumbres Observatory (FLOYDS, Sinistro), Magellan:Baade (FIRE), Magellan:Clay (MIKE), Gemini:South (FLAMINGOS-2), CTIO:PROMPT, SAAO:SALT (RSS), Swift (UVOT), VLT:Yepun (MUSE), VLT (FORS2), Gemini:South (GMOS)}

\software{Astropy \citep{astropy13,astropy18}, 
          lcogtsnpipe \citep{Valenti2016}, 
          SNID \citep{Blondin2007},
          SciPy (https://www.scipy.org),
          NumPy (https://numpy.org),
          PYRAF,
          HOTPANTS \citep{Becker2015},
          Matplotlib \citep{Hunter2007},
          Pandas \citep{mckinney-proc-scipy-2010}
          }

\bibliography{2018cuf}{}

\begin{thebibliography}{}
\expandafter\ifx\csname natexlab\endcsname\relax\def\natexlab#1{#1}\fi
\providecommand{\url}[1]{\href{#1}{#1}}
\providecommand{\dodoi}[1]{doi:~\href{http://doi.org/#1}{\nolinkurl{#1}}}
\providecommand{\doeprint}[1]{\href{http://ascl.net/#1}{\nolinkurl{http://ascl.net/#1}}}
\providecommand{\doarXiv}[1]{\href{https://arxiv.org/abs/#1}{\nolinkurl{https://arxiv.org/abs/#1}}}

\bibitem[{{Anderson}(2019)}]{Anderson2019}
{Anderson}, J.~P. 2019, \aap, 628, A7, \dodoi{10.1051/0004-6361/201935027}

\bibitem[{{Anderson} {et~al.}(2014){Anderson}, {Gonz{\'a}lez-Gait{\'a}n},
  {Hamuy}, {Guti{\'e}rrez}, {Stritzinger}, {Olivares E.}, {Phillips},
  {Schulze}, {Antezana}, {Bolt}, {Campillay}, {Castell{\'o}n}, {Contreras}, {de
  Jaeger}, {Folatelli}, {F{\"o}rster}, {Freedman}, {Gonz{\'a}lez}, {Hsiao},
  {Krzemi{\'n}ski}, {Krisciunas}, {Maza}, {McCarthy}, {Morrell}, {Persson},
  {Roth}, {Salgado}, {Suntzeff}, \& {Thomas-Osip}}]{Anderson2014}
{Anderson}, J.~P., {Gonz{\'a}lez-Gait{\'a}n}, S., {Hamuy}, M., {et~al.} 2014,
  \apj, 786, 67, \dodoi{10.1088/0004-637X/786/1/67}

\bibitem[{{Anderson} {et~al.}(2018){Anderson}, {Dessart}, {Guti{\'e}rrez},
  {Kr{\"u}hler}, {Galbany}, {Jerkstrand}, {Smartt}, {Contreras}, {Morrell},
  {Phillips}, {Stritzinger}, {Hsiao}, {Gonz{\'a}lez-Gait{\'a}n}, {Agliozzo},
  {Castell{\'o}n}, {Chambers}, {Chen}, {Flewelling}, {Gonzalez},
  {Hosseinzadeh}, {Huber}, {Fraser}, {Inserra}, {Kankare}, {Mattila},
  {Magnier}, {Maguire}, {Lowe}, {Sollerman}, {Sullivan}, {Young}, \&
  {Valenti}}]{anderson2018}
{Anderson}, J.~P., {Dessart}, L., {Guti{\'e}rrez}, C.~P., {et~al.} 2018, Nature
  Astronomy, 2, 574, \dodoi{10.1038/s41550-018-0458-4}

\bibitem[{{Appenzeller} {et~al.}(1998){Appenzeller}, {Fricke}, {F{\"u}rtig},
  {G{\"a}ssler}, {H{\"a}fner}, {Harke}, {Hess}, {Hummel}, {J{\"u}rgens},
  {Kudritzki}, {Mantel}, {Meisl}, {Muschielok}, {Nicklas}, {Rupprecht},
  {Seifert}, {Stahl}, {Szeifert}, \& {Tarantik}}]{Appenzeller1998}
{Appenzeller}, I., {Fricke}, K., {F{\"u}rtig}, W., {et~al.} 1998, The
  Messenger, 94, 1

\bibitem[{{Arcavi} {et~al.}(2017){Arcavi}, {Hosseinzadeh}, {Brown}, {Smartt},
  {Valenti}, {Tartaglia}, {Piro}, {Sanchez}, {Nicholls}, {Monard}, {Howell},
  {McCully}, {Sand}, {Tonry}, {Denneau}, {Stalder}, {Heinze}, {Rest}, {Smith},
  \& {Bishop}}]{Arcavi17}
{Arcavi}, I., {Hosseinzadeh}, G., {Brown}, P.~J., {et~al.} 2017, \apjl, 837,
  L2, \dodoi{10.3847/2041-8213/aa5be1}

\bibitem[{{Asplund} {et~al.}(2009){Asplund}, {Grevesse}, {Sauval}, \&
  {Scott}}]{Asplund2009}
{Asplund}, M., {Grevesse}, N., {Sauval}, A.~J., \& {Scott}, P. 2009, \araa, 47,
  481, \dodoi{10.1146/annurev.astro.46.060407.145222}

\bibitem[{{Astropy Collaboration} {et~al.}(2013){Astropy Collaboration},
  {Robitaille}, {Tollerud}, {Greenfield}, {Droettboom}, {Bray}, {Aldcroft},
  {Davis}, {Ginsburg}, {Price-Whelan}, {Kerzendorf}, {Conley}, {Crighton},
  {Barbary}, {Muna}, {Ferguson}, {Grollier}, {Parikh}, {Nair}, {Unther},
  {Deil}, {Woillez}, {Conseil}, {Kramer}, {Turner}, {Singer}, {Fox}, {Weaver},
  {Zabalza}, {Edwards}, {Azalee Bostroem}, {Burke}, {Casey}, {Crawford},
  {Dencheva}, {Ely}, {Jenness}, {Labrie}, {Lim}, {Pierfederici}, {Pontzen},
  {Ptak}, {Refsdal}, {Servillat}, \& {Streicher}}]{astropy13}
{Astropy Collaboration}, {Robitaille}, T.~P., {Tollerud}, E.~J., {et~al.} 2013,
  \aap, 558, A33, \dodoi{10.1051/0004-6361/201322068}

\bibitem[{{Astropy Collaboration} {et~al.}(2018){Astropy Collaboration},
  {Price-Whelan}, {Sip{\H o}cz}, {G{\"u}nther}, {Lim}, {Crawford}, {Conseil},
  {Shupe}, {Craig}, {Dencheva}, {Ginsburg}, {VanderPlas}, {Bradley},
  {P{\'e}rez-Su{\'a}rez}, {de Val-Borro}, {Aldcroft}, {Cruz}, {Robitaille},
  {Tollerud}, {Ardelean}, {Babej}, {Bach}, {Bachetti}, {Bakanov}, {Bamford},
  {Barentsen}, {Barmby}, {Baumbach}, {Berry}, {Biscani}, {Boquien}, {Bostroem},
  {Bouma}, {Brammer}, {Bray}, {Breytenbach}, {Buddelmeijer}, {Burke},
  {Calderone}, {Cano Rodr{\'{\i}}guez}, {Cara}, {Cardoso}, {Cheedella},
  {Copin}, {Corrales}, {Crichton}, {D'Avella}, {Deil}, {Depagne}, {Dietrich},
  {Donath}, {Droettboom}, {Earl}, {Erben}, {Fabbro}, {Ferreira}, {Finethy},
  {Fox}, {Garrison}, {Gibbons}, {Goldstein}, {Gommers}, {Greco}, {Greenfield},
  {Groener}, {Grollier}, {Hagen}, {Hirst}, {Homeier}, {Horton}, {Hosseinzadeh},
  {Hu}, {Hunkeler}, {Ivezi{\'c}}, {Jain}, {Jenness}, {Kanarek}, {Kendrew},
  {Kern}, {Kerzendorf}, {Khvalko}, {King}, {Kirkby}, {Kulkarni}, {Kumar},
  {Lee}, {Lenz}, {Littlefair}, {Ma}, {Macleod}, {Mastropietro}, {McCully},
  {Montagnac}, {Morris}, {Mueller}, {Mumford}, {Muna}, {Murphy}, {Nelson},
  {Nguyen}, {Ninan}, {N{\"o}the}, {Ogaz}, {Oh}, {Parejko}, {Parley}, {Pascual},
  {Patil}, {Patil}, {Plunkett}, {Prochaska}, {Rastogi}, {Reddy Janga},
  {Sabater}, {Sakurikar}, {Seifert}, {Sherbert}, {Sherwood-Taylor}, {Shih},
  {Sick}, {Silbiger}, {Singanamalla}, {Singer}, {Sladen}, {Sooley},
  {Sornarajah}, {Streicher}, {Teuben}, {Thomas}, {Tremblay}, {Turner},
  {Terr{\'o}n}, {van Kerkwijk}, {de la Vega}, {Watkins}, {Weaver}, {Whitmore},
  {Woillez}, {Zabalza}, \& {Astropy Contributors}}]{astropy18}
{Astropy Collaboration}, {Price-Whelan}, A.~M., {Sip{\H o}cz}, B.~M., {et~al.}
  2018, \aj, 156, 123, \dodoi{10.3847/1538-3881/aabc4f}

\bibitem[{{Baade}(1926)}]{Baade1926}
{Baade}, W. 1926, Astronomische Nachrichten, 228, 359,
  \dodoi{10.1002/asna.19262282003}

\bibitem[{{Bacon} {et~al.}(2010){Bacon}, {Accardo}, {Adjali}, {Anwand},
  {Bauer}, {Biswas}, {Blaizot}, {Boudon}, {Brau-Nogue}, {Brinchmann},
  {Caillier}, {Capoani}, {Carollo}, {Contini}, {Couderc}, {Daguis{\'e}},
  {Deiries}, {Delabre}, {Dreizler}, {Dubois}, {Dupieux}, {Dupuy}, {Emsellem},
  {Fechner}, {Fleischmann}, {Fran{\c{c}}ois}, {Gallou}, {Gharsa}, {Glindemann},
  {Gojak}, {Guiderdoni}, {Hansali}, {Hahn}, {Jarno}, {Kelz}, {Koehler},
  {Kosmalski}, {Laurent}, {Le Floch}, {Lilly}, {Lizon}, {Loupias}, {Manescau},
  {Monstein}, {Nicklas}, {Olaya}, {Pares}, {Pasquini}, {P{\'e}contal-Rousset},
  {Pell{\'o}}, {Petit}, {Popow}, {Reiss}, {Remillieux}, {Renault}, {Roth},
  {Rupprecht}, {Serre}, {Schaye}, {Soucail}, {Steinmetz}, {Streicher}, {Stuik},
  {Valentin}, {Vernet}, {Weilbacher}, {Wisotzki}, \& {Yerle}}]{Bacon2010}
{Bacon}, R., {Accardo}, M., {Adjali}, L., {et~al.} 2010, in Society of
  Photo-Optical Instrumentation Engineers (SPIE) Conference Series, Vol. 7735,
  Ground-based and Airborne Instrumentation for Astronomy III, 773508,
  \dodoi{10.1117/12.856027}

\bibitem[{{Becker}(2015)}]{Becker2015}
{Becker}, A. 2015, {HOTPANTS: High Order Transform of PSF ANd Template
  Subtraction}.
\newblock \doeprint{1504.004}

\bibitem[{{Bellm} {et~al.}(2019){Bellm}, {Kulkarni}, {Graham}, {Dekany},
  {Smith}, {Riddle}, {Masci}, {Helou}, {Prince}, {Adams}, {Barbarino},
  {Barlow}, {Bauer}, {Beck}, {Belicki}, {Biswas}, {Blagorodnova}, {Bodewits},
  {Bolin}, {Brinnel}, {Brooke}, {Bue}, {Bulla}, {Burruss}, {Cenko}, {Chang},
  {Connolly}, {Coughlin}, {Cromer}, {Cunningham}, {De}, {Delacroix}, {Desai},
  {Duev}, {Eadie}, {Farnham}, {Feeney}, {Feindt}, {Flynn}, {Franckowiak},
  {Frederick}, {Fremling}, {Gal-Yam}, {Gezari}, {Giomi}, {Goldstein},
  {Golkhou}, {Goobar}, {Groom}, {Hacopians}, {Hale}, {Henning}, {Ho}, {Hover},
  {Howell}, {Hung}, {Huppenkothen}, {Imel}, {Ip}, {Ivezi{\'c}}, {Jackson},
  {Jones}, {Juric}, {Kasliwal}, {Kaspi}, {Kaye}, {Kelley}, {Kowalski},
  {Kramer}, {Kupfer}, {Landry}, {Laher}, {Lee}, {Lin}, {Lin}, {Lunnan},
  {Giomi}, {Mahabal}, {Mao}, {Miller}, {Monkewitz}, {Murphy}, {Ngeow},
  {Nordin}, {Nugent}, {Ofek}, {Patterson}, {Penprase}, {Porter}, {Rauch},
  {Rebbapragada}, {Reiley}, {Rigault}, {Rodriguez}, {van Roestel}, {Rusholme},
  {van Santen}, {Schulze}, {Shupe}, {Singer}, {Soumagnac}, {Stein}, {Surace},
  {Sollerman}, {Szkody}, {Taddia}, {Terek}, {Van Sistine}, {van Velzen},
  {Vestrand}, {Walters}, {Ward}, {Ye}, {Yu}, {Yan}, \& {Zolkower}}]{Bellm2019}
{Bellm}, E.~C., {Kulkarni}, S.~R., {Graham}, M.~J., {et~al.} 2019, \pasp, 131,
  018002, \dodoi{10.1088/1538-3873/aaecbe}

\bibitem[{{Bernstein} {et~al.}(2003){Bernstein}, {Shectman}, {Gunnels},
  {Mochnacki}, \& {Athey}}]{Bernstei2003}
{Bernstein}, R., {Shectman}, S.~A., {Gunnels}, S.~M., {Mochnacki}, S., \&
  {Athey}, A.~E. 2003, in Society of Photo-Optical Instrumentation Engineers
  (SPIE) Conference Series, Vol. 4841, \procspie, ed. M.~{Iye} \& A.~F.~M.
  {Moorwood}, 1694--1704, \dodoi{10.1117/12.461502}

\bibitem[{{Bersten} {et~al.}(2011){Bersten}, {Benvenuto}, \&
  {Hamuy}}]{Bersten2011}
{Bersten}, M.~C., {Benvenuto}, O., \& {Hamuy}, M. 2011, \apj, 729, 61,
  \dodoi{10.1088/0004-637X/729/1/61}

\bibitem[{{Bersten} {et~al.}(2018){Bersten}, {Folatelli}, {Garc{\'\i}a}, {van
  Dyk}, {Benvenuto}, {Orellana}, {Buso}, {S{\'a}nchez}, {Tanaka}, {Maeda},
  {Filippenko}, {Zheng}, {Brink}, {Cenko}, {de Jaeger}, {Kumar}, {Moriya},
  {Nomoto}, {Perley}, {Shivvers}, \& {Smith}}]{Bersten18}
{Bersten}, M.~C., {Folatelli}, G., {Garc{\'\i}a}, F., {et~al.} 2018, \nat, 554,
  497, \dodoi{10.1038/nature25151}

\bibitem[{{Blondin} \& {Tonry}(2007)}]{Blondin2007}
{Blondin}, S., \& {Tonry}, J.~L. 2007, \apj, 666, 1024, \dodoi{10.1086/520494}

\bibitem[{{Bose} {et~al.}(2020){Bose}, {Dong}, {Kochanek}, {Stritzinger},
  {Ashall}, {Benetti}, {Falco}, {Filippenko}, {Pastorello}, {Prieto}, {Somero},
  {Sukhbold}, {Zhang}, {Auchettl}, {Brink}, {Brown}, {Chen}, {Fiore}, {Grupe},
  {Holoien}, {Lundqvist}, {Mattila}, {Mutel}, {Pooley}, {Post}, {Reddy},
  {Reynolds}, {Shappee}, {Stanek}, {Thompson}, {Villanueva}, \&
  {Zheng}}]{bose2020}
{Bose}, S., {Dong}, S., {Kochanek}, C.~S., {et~al.} 2020, arXiv e-prints,
  arXiv:2007.00008.
\newblock \doarXiv{2007.00008}

\bibitem[{{Bostroem} {et~al.}(2019){Bostroem}, {Valenti}, {Horesh}, {Morozova},
  {Kuin}, {Wyatt}, {Jerkstrand}, {Sand}, {Lundquist}, {Smith}, {Sullivan},
  {Hosseinzadeh}, {Arcavi}, {Callis}, {Cartier}, {Gal-Yam}, {Galbany},
  {Guti{\'e}rrez}, {Howell}, {Inserra}, {Kankare}, {L{\'o}pez}, {McCully},
  {Pignata}, {Piro}, {Rodr{\'\i}guez}, {Smartt}, {Smith}, {Yaron}, \&
  {Young}}]{Bostroem2019}
{Bostroem}, K.~A., {Valenti}, S., {Horesh}, A., {et~al.} 2019, \mnras, 485,
  5120, \dodoi{10.1093/mnras/stz570}

\bibitem[{{Bostroem} {et~al.}(2020){Bostroem}, {Valenti}, {Sand}, {Andrews},
  {Van Dyk}, {Galbany}, {Pooley}, {Amaro}, {Smith}, {Yang}, {Anupama},
  {Arcavi}, {Baron}, {Brown}, {Burke}, {Cartier}, {Hiramatsu}, {Dastidar},
  {DerKacy}, {Dong}, {Egami}, {Ertel}, {Filippenko}, {Fox}, {Haislip},
  {Hosseinzadeh}, {Howell}, {Gangopadhyay}, {Jha}, {Kouprianov}, {Kumar},
  {Lundquist}, {Milisavljevic}, {McCully}, {Milne}, {Misra}, {Reichart},
  {Sahu}, {Sai}, {Singh}, {Smith}, {Vinko}, {Wang}, {Wang}, {Wheeler},
  {Williams}, {Wyatt}, {Zhang}, \& {Zhang}}]{Bostroem2020}
{Bostroem}, K.~A., {Valenti}, S., {Sand}, D.~J., {et~al.} 2020, \apj, 895, 31,
  \dodoi{10.3847/1538-4357/ab8945}

\bibitem[{{Breeveld} {et~al.}(2011){Breeveld}, {Landsman}, {Holland}, {Roming},
  {Kuin}, \& {Page}}]{Breeveld2011}
{Breeveld}, A.~A., {Landsman}, W., {Holland}, S.~T., {et~al.} 2011, in American
  Institute of Physics Conference Series, Vol. 1358, American Institute of
  Physics Conference Series, ed. J.~E. {McEnery}, J.~L. {Racusin}, \&
  N.~{Gehrels}, 373--376, \dodoi{10.1063/1.3621807}

\bibitem[{{Brown} {et~al.}(2014){Brown}, {Breeveld}, {Holland}, {Kuin}, \&
  {Pritchard}}]{Brown2014}
{Brown}, P.~J., {Breeveld}, A.~A., {Holland}, S., {Kuin}, P., \& {Pritchard},
  T. 2014, \apss, 354, 89, \dodoi{10.1007/s10509-014-2059-8}

\bibitem[{{Brown} {et~al.}(2009){Brown}, {Holland}, {Immler}, {Milne},
  {Roming}, {Gehrels}, {Nousek}, {Panagia}, {Still}, \& {Vand en
  Berk}}]{Brown2009}
{Brown}, P.~J., {Holland}, S.~T., {Immler}, S., {et~al.} 2009, \aj, 137, 4517,
  \dodoi{10.1088/0004-6256/137/5/4517}

\bibitem[{{Brown} {et~al.}(2013){Brown}, {Baliber}, {Bianco}, {Bowman},
  {Burleson}, {Conway}, {Crellin}, {Depagne}, {De Vera}, {Dilday}, {Dragomir},
  {Dubberley}, {Eastman}, {Elphick}, {Falarski}, {Foale}, {Ford}, {Fulton},
  {Garza}, {Gomez}, {Graham}, {Greene}, {Haldeman}, {Hawkins}, {Haworth},
  {Haynes}, {Hidas}, {Hjelstrom}, {Howell}, {Hygelund}, {Lister}, {Lobdill},
  {Martinez}, {Mullins}, {Norbury}, {Parrent}, {Paulson}, {Petry}, {Pickles},
  {Posner}, {Rosing}, {Ross}, {Sand}, {Saunders}, {Shobbrook}, {Shporer},
  {Street}, {Thomas}, {Tsapras}, {Tufts}, {Valenti}, {Vander Horst}, {Walker},
  {White}, \& {Willis}}]{Brown2013}
{Brown}, T.~M., {Baliber}, N., {Bianco}, F.~B., {et~al.} 2013, \pasp, 125,
  1031, \dodoi{10.1086/673168}

\bibitem[{{Cardelli} {et~al.}(1989){Cardelli}, {Clayton}, \&
  {Mathis}}]{Cardelli1989}
{Cardelli}, J.~A., {Clayton}, G.~C., \& {Mathis}, J.~S. 1989, \apj, 345, 245,
  \dodoi{10.1086/167900}

\bibitem[{{Chevalier} \& {Fransson}(1994)}]{Chevalier1994}
{Chevalier}, R.~A., \& {Fransson}, C. 1994, \apj, 420, 268,
  \dodoi{10.1086/173557}

\bibitem[{{Chugai} {et~al.}(2007){Chugai}, {Chevalier}, \&
  {Utrobin}}]{Chugai2007}
{Chugai}, N.~N., {Chevalier}, R.~A., \& {Utrobin}, V.~P. 2007, \apj, 662, 1136,
  \dodoi{10.1086/518160}

\bibitem[{{Czerny} {et~al.}(2018){Czerny}, {Beaton}, {Bejger}, {Cackett},
  {Dall'Ora}, {Holanda}, {Jensen}, {Jha}, {Lusso}, {Minezaki}, {Risaliti},
  {Salaris}, {Toonen}, \& {Yoshii}}]{Czerny2018}
{Czerny}, B., {Beaton}, R., {Bejger}, M., {et~al.} 2018, \ssr, 214, 32,
  \dodoi{10.1007/s11214-018-0466-9}

\bibitem[{{Davies} \& {Beasor}(2018)}]{Davies2018}
{Davies}, B., \& {Beasor}, E.~R. 2018, \mnras, 474, 2116,
  \dodoi{10.1093/mnras/stx2734}

\bibitem[{{Davies} \& {Beasor}(2020)}]{Davies2020}
---. 2020, \mnras, 496, L142, \dodoi{10.1093/mnrasl/slaa102}

\bibitem[{{Davis} {et~al.}(2019){Davis}, {Hsiao}, {Ashall}, {Hoeflich},
  {Phillips}, {Marion}, {Kirshner}, {Morrell}, {Sand}, {Burns}, {Contreras},
  {Stritzinger}, {Anderson}, {Baron}, {Diamond}, {Guti{\'e}rrez}, {Hamuy},
  {Holmbo}, {Kasliwal}, {Krisciunas}, {Kumar}, {Lu}, {Pessi}, {Piro}, {Prieto},
  {Shahbandeh}, \& {Suntzeff}}]{Davis2019}
{Davis}, S., {Hsiao}, E.~Y., {Ashall}, C., {et~al.} 2019, \apj, 887, 4,
  \dodoi{10.3847/1538-4357/ab4c40}

\bibitem[{{de Jaeger} {et~al.}(2018){de Jaeger}, {Anderson}, {Galbany},
  {Gonz{\'a}lez-Gait{\'a}n}, {Hamuy}, {Phillips}, {Stritzinger}, {Contreras},
  {Folatelli}, {Guti{\'e}rrez}, {Hsiao}, {Morrell}, {Suntzeff}, {Dessart}, \&
  {Filippenko}}]{Jaeger2018}
{de Jaeger}, T., {Anderson}, J.~P., {Galbany}, L., {et~al.} 2018, \mnras, 476,
  4592, \dodoi{10.1093/mnras/sty508}

\bibitem[{{Dessart} \& {Hillier}(2005)}]{Dessart2005}
{Dessart}, L., \& {Hillier}, D.~J. 2005, \aap, 439, 671,
  \dodoi{10.1051/0004-6361:20053217}

\bibitem[{{Eikenberry} {et~al.}(2006){Eikenberry}, {Elston}, {Raines},
  {Julian}, {Hanna}, {Hon}, {Julian}, {Bandyopadhyay}, {Bennett}, {Bessoff},
  {Branch}, {Corley}, {Eriksen}, {Frommeyer}, {Gonzalez}, {Herlevich},
  {Marin-Franch}, {Marti}, {Murphey}, {Rashkin}, {Warner}, {Leckie},
  {Gardhouse}, {Fletcher}, {Dunn}, {Wooff}, \& {Hardy}}]{Eikenberry2006}
{Eikenberry}, S., {Elston}, R., {Raines}, S.~N., {et~al.} 2006, Society of
  Photo-Optical Instrumentation Engineers (SPIE) Conference Series, Vol. 6269,
  {FLAMINGOS-2: the facility near-infrared wide-field imager and multi-object
  spectrograph for Gemini}, 626917, \dodoi{10.1117/12.672095}

\bibitem[{{Ekstr{\"o}m} {et~al.}(2012){Ekstr{\"o}m}, {Georgy}, {Eggenberger},
  {Meynet}, {Mowlavi}, {Wyttenbach}, {Granada}, {Decressin}, {Hirschi},
  {Frischknecht}, {Charbonnel}, \& {Maeder}}]{Ekstrom2012}
{Ekstr{\"o}m}, S., {Georgy}, C., {Eggenberger}, P., {et~al.} 2012, \aap, 537,
  A146, \dodoi{10.1051/0004-6361/201117751}

\bibitem[{{Faran} {et~al.}(2014){Faran}, {Poznanski}, {Filippenko}, {Chornock},
  {Foley}, {Ganeshalingam}, {Leonard}, {Li}, {Modjaz}, {Nakar}, {Serduke}, \&
  {Silverman}}]{Faran2014}
{Faran}, T., {Poznanski}, D., {Filippenko}, A.~V., {et~al.} 2014, \mnras, 442,
  844, \dodoi{10.1093/mnras/stu955}

\bibitem[{{Foreman-Mackey} {et~al.}(2013){Foreman-Mackey}, {Hogg}, {Lang}, \&
  {Goodman}}]{Foreman2013}
{Foreman-Mackey}, D., {Hogg}, D.~W., {Lang}, D., \& {Goodman}, J. 2013, \pasp,
  125, 306, \dodoi{10.1086/670067}

\bibitem[{{Freudling} {et~al.}(2013){Freudling}, {Romaniello}, {Bramich},
  {Ballester}, {Forchi}, {Garc{\'\i}a-Dabl{\'o}}, {Moehler}, \&
  {Neeser}}]{Freudling2013}
{Freudling}, W., {Romaniello}, M., {Bramich}, D.~M., {et~al.} 2013, \aap, 559,
  A96, \dodoi{10.1051/0004-6361/201322494}

\bibitem[{{Fuller}(2017)}]{Fuller2017}
{Fuller}, J. 2017, \mnras, 470, 1642, \dodoi{10.1093/mnras/stx1314}

\bibitem[{{Galbany} {et~al.}(2016{\natexlab{a}}){Galbany}, {Hamuy}, {Phillips},
  {Suntzeff}, {Maza}, {de Jaeger}, {Moraga}, {Gonz{\'a}lez-Gait{\'a}n},
  {Krisciunas}, {Morrell}, {Thomas-Osip}, {Krzeminski}, {Gonz{\'a}lez},
  {Antezana}, {Wishnjewski}, {McCarthy}, {Anderson}, {Guti{\'e}rrez},
  {Stritzinger}, {Folatelli}, {Anguita}, {Galaz}, {Green}, {Impey}, {Kim},
  {Kirhakos}, {Malkan}, {Mulchaey}, {Phillips}, {Pizzella}, {Prosser},
  {Schmidt}, {Schommer}, {Sherry}, {Strolger}, {Wells}, \&
  {Williger}}]{Galbany2016AJ}
{Galbany}, L., {Hamuy}, M., {Phillips}, M.~M., {et~al.} 2016{\natexlab{a}},
  \aj, 151, 33, \dodoi{10.3847/0004-6256/151/2/33}

\bibitem[{{Galbany} {et~al.}(2016{\natexlab{b}}){Galbany}, {Anderson},
  {Rosales-Ortega}, {Kuncarayakti}, {Kr{\"u}hler}, {S{\'a}nchez},
  {Falc{\'o}n-Barroso}, {P{\'e}rez}, {Maureira}, {Hamuy},
  {Gonz{\'a}lez-Gait{\'a}n}, {F{\"o}rster}, \& {Moral}}]{Galbany2016}
{Galbany}, L., {Anderson}, J.~P., {Rosales-Ortega}, F.~F., {et~al.}
  2016{\natexlab{b}}, \mnras, 455, 4087, \dodoi{10.1093/mnras/stv2620}

\bibitem[{{Galbany} {et~al.}(2018){Galbany}, {Anderson}, {S{\'a}nchez},
  {Kuncarayakti}, {Pedraz}, {Gonz{\'a}lez-Gait{\'a}n}, {Stanishev},
  {Dom{\'\i}nguez}, {Moreno-Raya}, {Wood-Vasey}, {Mour{\~a}o}, {Ponder},
  {Badenes}, {Moll{\'a}}, {L{\'o}pez-S{\'a}nchez}, {Rosales-Ortega},
  {V{\'\i}lchez}, {Garc{\'\i}a-Benito}, \& {Marino}}]{Galbany2018}
{Galbany}, L., {Anderson}, J.~P., {S{\'a}nchez}, S.~F., {et~al.} 2018, \apj,
  855, 107, \dodoi{10.3847/1538-4357/aaaf20}

\bibitem[{{Gall} {et~al.}(2014){Gall}, {Hjorth}, {Watson}, {Dwek}, {Maund},
  {Fox}, {Leloudas}, {Malesani}, \& {Day-Jones}}]{gall14}
{Gall}, C., {Hjorth}, J., {Watson}, D., {et~al.} 2014, \nat, 511, 326,
  \dodoi{10.1038/nature13558}

\bibitem[{{Gehrels} {et~al.}(2004){Gehrels}, {Chincarini}, {Giommi}, {Mason},
  {Nousek}, {Wells}, {White}, {Barthelmy}, {Burrows}, {Cominsky}, {Hurley},
  {Marshall}, {M{\'e}sz{\'a}ros}, {Roming}, {Angelini}, {Barbier}, {Belloni},
  {Campana}, {Caraveo}, {Chester}, {Citterio}, {Cline}, {Cropper}, {Cummings},
  {Dean}, {Feigelson}, {Fenimore}, {Frail}, {Fruchter}, {Garmire}, {Gendreau},
  {Ghisellini}, {Greiner}, {Hill}, {Hunsberger}, {Krimm}, {Kulkarni}, {Kumar},
  {Lebrun}, {Lloyd-Ronning}, {Markwardt}, {Mattson}, {Mushotzky}, {Norris},
  {Osborne}, {Paczynski}, {Palmer}, {Park}, {Parsons}, {Paul}, {Rees},
  {Reynolds}, {Rhoads}, {Sasseen}, {Schaefer}, {Short}, {Smale}, {Smith},
  {Stella}, {Tagliaferri}, {Takahashi}, {Tashiro}, {Townsley}, {Tueller},
  {Turner}, {Vietri}, {Voges}, {Ward}, {Willingale}, {Zerbi}, \&
  {Zhang}}]{Gehrels2004}
{Gehrels}, N., {Chincarini}, G., {Giommi}, P., {et~al.} 2004, \apj, 611, 1005,
  \dodoi{10.1086/422091}

\bibitem[{Gimeno {et~al.}(2016)Gimeno, Roth, Chiboucas, Hibon, Boucher, White,
  Rippa, Labrie, Turner, Hanna, Lazo, Pérez, Rogers, Rojas, Placco, \&
  Murowinski}]{10.1117/12.2233883}
Gimeno, G., Roth, K., Chiboucas, K., {et~al.} 2016, in Ground-based and
  Airborne Instrumentation for Astronomy VI, ed. C.~J. Evans, L.~Simard, \&
  H.~Takami, Vol. 9908, International Society for Optics and Photonics (SPIE),
  872 -- 885, \dodoi{10.1117/12.2233883}

\bibitem[{{Goldberg} \& {Bildsten}(2020)}]{Goldberg2020}
{Goldberg}, J.~A., \& {Bildsten}, L. 2020, \apjl, 895, L45,
  \dodoi{10.3847/2041-8213/ab9300}

\bibitem[{{Goldberg} {et~al.}(2019){Goldberg}, {Bildsten}, \&
  {Paxton}}]{Goldberg2019}
{Goldberg}, J.~A., {Bildsten}, L., \& {Paxton}, B. 2019, \apj, 879, 3,
  \dodoi{10.3847/1538-4357/ab22b6}

\bibitem[{{Guti{\'e}rrez} {et~al.}(2017){Guti{\'e}rrez}, {Anderson}, {Hamuy},
  {Morrell}, {Gonz{\'a}lez-Gaitan}, {Stritzinger}, {Phillips}, {Galbany},
  {Folatelli}, {Dessart}, {Contreras}, {Della Valle}, {Freedman}, {Hsiao},
  {Krisciunas}, {Madore}, {Maza}, {Suntzeff}, {Prieto}, {Gonz{\'a}lez},
  {Cappellaro}, {Navarrete}, {Pizzella}, {Ruiz}, {Smith}, \&
  {Turatto}}]{Gutierrez2017a}
{Guti{\'e}rrez}, C.~P., {Anderson}, J.~P., {Hamuy}, M., {et~al.} 2017, \apj,
  850, 89, \dodoi{10.3847/1538-4357/aa8f52}

\bibitem[{{Hamuy} {et~al.}(2001){Hamuy}, {Pinto}, {Maza}, {Suntzeff},
  {Phillips}, {Eastman}, {Smith}, {Corbally}, {Burstein}, {Li}, {Ivanov},
  {Moro-Martin}, {Strolger}, {de Souza}, {dos Anjos}, {Green}, {Pickering},
  {Gonz{\'a}lez}, {Antezana}, {Wischnjewsky}, {Galaz}, {Roth}, {Persson}, \&
  {Schommer}}]{Hamuy2001}
{Hamuy}, M., {Pinto}, P.~A., {Maza}, J., {et~al.} 2001, \apj, 558, 615,
  \dodoi{10.1086/322450}

\bibitem[{{Heger} {et~al.}(2003){Heger}, {Fryer}, {Woosley}, {Langer}, \&
  {Hartmann}}]{Heger2003}
{Heger}, A., {Fryer}, C.~L., {Woosley}, S.~E., {Langer}, N., \& {Hartmann},
  D.~H. 2003, \apj, 591, 288, \dodoi{10.1086/375341}

\bibitem[{{Hook} {et~al.}(2004){Hook}, {J{\o}rgensen}, {Allington-Smith},
  {Davies}, {Metcalfe}, {Murowinski}, \& {Crampton}}]{Hook2004}
{Hook}, I.~M., {J{\o}rgensen}, I., {Allington-Smith}, J.~R., {et~al.} 2004,
  \pasp, 116, 425, \dodoi{10.1086/383624}

\bibitem[{{Horiuchi} {et~al.}(2014){Horiuchi}, {Nakamura}, {Takiwaki},
  {Kotake}, \& {Tanaka}}]{Horiuchi2014}
{Horiuchi}, S., {Nakamura}, K., {Takiwaki}, T., {Kotake}, K., \& {Tanaka}, M.
  2014, \mnras, 445, L99, \dodoi{10.1093/mnrasl/slu146}

\bibitem[{Hosseinzadeh(2020)}]{griffin_hosseinzadeh2020}
Hosseinzadeh, G. 2020, griffin-h/lightcurve\_fitting v0.1.0, v0.1.0,  Zenodo,
  \dodoi{10.5281/zenodo.3908580}

\bibitem[{{Hosseinzadeh} {et~al.}(2018){Hosseinzadeh}, {Valenti}, {McCully},
  {Howell}, {Arcavi}, {Jerkstrand}, {Guevel}, {Tartaglia}, {Rui}, {Mo}, {Wang},
  {Huang}, {Song}, {Zhang}, \& {Itagaki}}]{Hosseinzadeh2018}
{Hosseinzadeh}, G., {Valenti}, S., {McCully}, C., {et~al.} 2018, \apj, 861, 63,
  \dodoi{10.3847/1538-4357/aac5f6}

\bibitem[{{Hsiao} {et~al.}(2019){Hsiao}, {Phillips}, {Marion}, {Kirshner},
  {Morrell}, {Sand}, {Burns}, {Contreras}, {Hoeflich}, {Stritzinger},
  {Valenti}, {Anderson}, {Ashall}, {Baltay}, {Baron}, {Banerjee}, {Davis},
  {Diamond}, {Folatelli}, {Freedman}, {F{\"o}rster}, {Galbany}, {Gall},
  {Gonz{\'a}lez-Gait{\'a}n}, {Goobar}, {Hamuy}, {Holmbo}, {Kasliwal},
  {Krisciunas}, {Kumar}, {Lidman}, {Lu}, {Nugent}, {Perlmutter}, {Persson},
  {Piro}, {Rabinowitz}, {Roth}, {Ryder}, {Schmidt}, {Shahbandeh}, {Suntzeff},
  {Taddia}, {Uddin}, \& {Wang}}]{Hsiao19}
{Hsiao}, E.~Y., {Phillips}, M.~M., {Marion}, G.~H., {et~al.} 2019, \pasp, 131,
  014002, \dodoi{10.1088/1538-3873/aae961}

\bibitem[{{Hunter}(2007)}]{Hunter2007}
{Hunter}, J.~D. 2007, Computing in Science and Engineering, 9, 90,
  \dodoi{10.1109/MCSE.2007.55}

\bibitem[{{Inserra} {et~al.}(2012){Inserra}, {Turatto}, {Pastorello}, {Pumo},
  {Baron}, {Benetti}, {Cappellaro}, {Taubenberger}, {Bufano}, {Elias-Rosa},
  {Zampieri}, {Harutyunyan}, {Moskvitin}, {Nissinen}, {Stanishev}, {Tsvetkov},
  {Hentunen}, {Komarova}, {Pavlyuk}, {Sokolov}, \& {Sokolova}}]{Inserra2012}
{Inserra}, C., {Turatto}, M., {Pastorello}, A., {et~al.} 2012, \mnras, 422,
  1122, \dodoi{10.1111/j.1365-2966.2012.20685.x}

\bibitem[{{Jerkstrand} {et~al.}(2018){Jerkstrand}, {Ertl}, {Janka},
  {M{\"u}ller}, {Sukhbold}, \& {Woosley}}]{Jerkstrand2018}
{Jerkstrand}, A., {Ertl}, T., {Janka}, H.~T., {et~al.} 2018, \mnras, 475, 277,
  \dodoi{10.1093/mnras/stx2877}

\bibitem[{{Jerkstrand} {et~al.}(2011){Jerkstrand}, {Fransson}, \&
  {Kozma}}]{Jerkstrand2011}
{Jerkstrand}, A., {Fransson}, C., \& {Kozma}, C. 2011, \aap, 530, A45,
  \dodoi{10.1051/0004-6361/201015937}

\bibitem[{{Jerkstrand} {et~al.}(2012){Jerkstrand}, {Fransson}, {Maguire},
  {Smartt}, {Ergon}, \& {Spyromilio}}]{Jerkstrand2012}
{Jerkstrand}, A., {Fransson}, C., {Maguire}, K., {et~al.} 2012, \aap, 546, A28,
  \dodoi{10.1051/0004-6361/201219528}

\bibitem[{{Jerkstrand} {et~al.}(2014){Jerkstrand}, {Smartt}, {Fraser},
  {Fransson}, {Sollerman}, {Taddia}, \& {Kotak}}]{Jerkstrand2014}
{Jerkstrand}, A., {Smartt}, S.~J., {Fraser}, M., {et~al.} 2014, \mnras, 439,
  3694, \dodoi{10.1093/mnras/stu221}

\bibitem[{{Jha}(2018)}]{Jha2018}
{Jha}, S. 2018, Transient Name Server Classification Report, 2018-884, 1

\bibitem[{{Jones} {et~al.}(2009){Jones}, {Hamuy}, {Lira}, {Maza},
  {Clocchiatti}, {Phillips}, {Morrell}, {Roth}, {Suntzeff}, {Matheson},
  {Filippenko}, {Foley}, \& {Leonard}}]{Jones2009}
{Jones}, M.~I., {Hamuy}, M., {Lira}, P., {et~al.} 2009, \apj, 696, 1176,
  \dodoi{10.1088/0004-637X/696/2/1176}

\bibitem[{{Kasen} \& {Woosley}(2009)}]{Kasen2009}
{Kasen}, D., \& {Woosley}, S.~E. 2009, \apj, 703, 2205,
  \dodoi{10.1088/0004-637X/703/2/2205}

\bibitem[{{Kennicutt}(1998)}]{Kennicutt1998}
{Kennicutt}, Robert~C., J. 1998, \apj, 498, 541, \dodoi{10.1086/305588}

\bibitem[{{Kirshner} \& {Kwan}(1974)}]{Kirshner1974}
{Kirshner}, R.~P., \& {Kwan}, J. 1974, \apj, 193, 27, \dodoi{10.1086/153123}

\bibitem[{{Kochanek} {et~al.}(2012){Kochanek}, {Khan}, \& {Dai}}]{Kochanek2012}
{Kochanek}, C.~S., {Khan}, R., \& {Dai}, X. 2012, \apj, 759, 20,
  \dodoi{10.1088/0004-637X/759/1/20}

\bibitem[{{Kochanek} {et~al.}(2017){Kochanek}, {Shappee}, {Stanek}, {Holoien},
  {Thompson}, {Prieto}, {Dong}, {Shields}, {Will}, {Britt}, {Perzanowski}, \&
  {Pojma{\'n}ski}}]{Kochanek2017}
{Kochanek}, C.~S., {Shappee}, B.~J., {Stanek}, K.~Z., {et~al.} 2017, \pasp,
  129, 104502, \dodoi{10.1088/1538-3873/aa80d9}

\bibitem[{{Leonard} {et~al.}(2002{\natexlab{a}}){Leonard}, {Filippenko}, {Li},
  {Matheson}, {Kirshner}, {Chornock}, {Van Dyk}, {Berlind}, {Calkins},
  {Challis}, {Garnavich}, {Jha}, \& {Mahdavi}}]{Leonard2002a}
{Leonard}, D.~C., {Filippenko}, A.~V., {Li}, W., {et~al.} 2002{\natexlab{a}},
  \aj, 124, 2490, \dodoi{10.1086/343771}

\bibitem[{{Leonard} {et~al.}(2002{\natexlab{b}}){Leonard}, {Filippenko},
  {Gates}, {Li}, {Eastman}, {Barth}, {Bus}, {Chornock}, {Coil}, {Frink},
  {Grady}, {Harris}, {Malkan}, {Matheson}, {Quirrenbach}, \&
  {Treffers}}]{Leonard2002}
{Leonard}, D.~C., {Filippenko}, A.~V., {Gates}, E.~L., {et~al.}
  2002{\natexlab{b}}, \pasp, 114, 35, \dodoi{10.1086/324785}

\bibitem[{{Li} {et~al.}(2011){Li}, {Leaman}, {Chornock}, {Filippenko},
  {Poznanski}, {Ganeshalingam}, {Wang}, {Modjaz}, {Jha}, {Foley}, \&
  {Smith}}]{Li2011}
{Li}, W., {Leaman}, J., {Chornock}, R., {et~al.} 2011, \mnras, 412, 1441,
  \dodoi{10.1111/j.1365-2966.2011.18160.x}

\bibitem[{{Lupton} {et~al.}(2005){Lupton}, {Juri{\'c}}, {Ivezi{\'c}}, {Brooks},
  {Schlegel}, {Finkbeiner}, {Padmanabhan}, {Bond}, {Rockosi}, {Knapp}, {Gunn},
  {Sumi}, \& {Schneider}}]{Lupton2005}
{Lupton}, R.~H., {Juri{\'c}}, M., {Ivezi{\'c}}, Z., {et~al.} 2005, in American
  Astronomical Society Meeting Abstracts, Vol. 207, 133.08

\bibitem[{{Martinez} \& {Bersten}(2019)}]{Martinez2019}
{Martinez}, L., \& {Bersten}, M.~C. 2019, \aap, 629, A124,
  \dodoi{10.1051/0004-6361/201834818}

\bibitem[{{Mathewson} {et~al.}(1992){Mathewson}, {Ford}, \&
  {Buchhorn}}]{Mathewson1992}
{Mathewson}, D.~S., {Ford}, V.~L., \& {Buchhorn}, M. 1992, \apjs, 81, 413,
  \dodoi{10.1086/191700}

\bibitem[{{Mauerhan} \& {Smith}(2012)}]{mauerhan12}
{Mauerhan}, J., \& {Smith}, N. 2012, \mnras, 424, 2659,
  \dodoi{10.1111/j.1365-2966.2012.21325.x}

\bibitem[{{Morozova} {et~al.}(2020){Morozova}, {Piro}, {Fuller}, \& {Van
  Dyk}}]{Morozova2020}
{Morozova}, V., {Piro}, A.~L., {Fuller}, J., \& {Van Dyk}, S.~D. 2020, \apjl,
  891, L32, \dodoi{10.3847/2041-8213/ab77c8}

\bibitem[{{Morozova} {et~al.}(2015){Morozova}, {Piro}, {Renzo}, {Ott},
  {Clausen}, {Couch}, {Ellis}, \& {Roberts}}]{Morozova2015}
{Morozova}, V., {Piro}, A.~L., {Renzo}, M., {et~al.} 2015, \apj, 814, 63,
  \dodoi{10.1088/0004-637X/814/1/63}

\bibitem[{{Morozova} {et~al.}(2017){Morozova}, {Piro}, \&
  {Valenti}}]{Morozova2017}
{Morozova}, V., {Piro}, A.~L., \& {Valenti}, S. 2017, \apj, 838, 28,
  \dodoi{10.3847/1538-4357/aa6251}

\bibitem[{{Morozova} {et~al.}(2018){Morozova}, {Piro}, \&
  {Valenti}}]{Morozova2018}
---. 2018, \apj, 858, 15, \dodoi{10.3847/1538-4357/aab9a6}

\bibitem[{{M{\"u}ller} {et~al.}(2017){M{\"u}ller}, {Prieto}, {Pejcha}, \&
  {Clocchiatti}}]{Muller2017}
{M{\"u}ller}, T., {Prieto}, J.~L., {Pejcha}, O., \& {Clocchiatti}, A. 2017,
  \apj, 841, 127, \dodoi{10.3847/1538-4357/aa72f1}

\bibitem[{{Munari} \& {Zwitter}(1997)}]{Munari1997}
{Munari}, U., \& {Zwitter}, T. 1997, \aap, 318, 269

\bibitem[{{Nakaoka} {et~al.}(2018){Nakaoka}, {Kawabata}, {Maeda}, {Tanaka},
  {Yamanaka}, {Moriya}, {Tominaga}, {Morokuma}, {Takaki}, {Kawabata},
  {Kawahara}, {Itoh}, {Shiki}, {Mori}, {Hirochi}, {Abe}, {Uemura}, {Yoshida},
  {Akitaya}, {Moritani}, {Ueno}, {Urano}, {Isogai}, {Hanayama}, \&
  {Nagayama}}]{Nakaoka2018}
{Nakaoka}, T., {Kawabata}, K.~S., {Maeda}, K., {et~al.} 2018, \apj, 859, 78,
  \dodoi{10.3847/1538-4357/aabee7}

\bibitem[{{Olivares E.} {et~al.}(2010){Olivares E.}, {Hamuy}, {Pignata},
  {Maza}, {Bersten}, {Phillips}, {Suntzeff}, {Filippenko}, {Morrel},
  {Kirshner}, \& {Matheson}}]{Olivares2010}
{Olivares E.}, F., {Hamuy}, M., {Pignata}, G., {et~al.} 2010, \apj, 715, 833,
  \dodoi{10.1088/0004-637X/715/2/833}

\bibitem[{{Osterbrock} \& {Ferland}(2006)}]{Osterbrock2006}
{Osterbrock}, D.~E., \& {Ferland}, G.~J. 2006, {Astrophysics of gaseous nebulae
  and active galactic nuclei}

\bibitem[{{Ouchi} \& {Maeda}(2019)}]{Ouchi2019}
{Ouchi}, R., \& {Maeda}, K. 2019, \apj, 877, 92,
  \dodoi{10.3847/1538-4357/ab1a37}

\bibitem[{{Pastorello} {et~al.}(2006){Pastorello}, {Sauer}, {Taubenberger},
  {Mazzali}, {Nomoto}, {Kawabata}, {Benetti}, {Elias-Rosa}, {Harutyunyan},
  {Navasardyan}, {Zampieri}, {Iijima}, {Botticella}, {di Rico}, {Del Principe},
  {Dolci}, {Gagliardi}, {Ragni}, \& {Valentini}}]{Pastorello2006}
{Pastorello}, A., {Sauer}, D., {Taubenberger}, S., {et~al.} 2006, \mnras, 370,
  1752, \dodoi{10.1111/j.1365-2966.2006.10587.x}

\bibitem[{{Paxton} {et~al.}(2018){Paxton}, {Schwab}, {Bauer}, {Bildsten},
  {Blinnikov}, {Duffell}, {Farmer}, {Goldberg}, {Marchant}, {Sorokina},
  {Thoul}, {Townsend}, \& {Timmes}}]{Paxton2018}
{Paxton}, B., {Schwab}, J., {Bauer}, E.~B., {et~al.} 2018, \apjs, 234, 34,
  \dodoi{10.3847/1538-4365/aaa5a8}

\bibitem[{{Pettini} \& {Pagel}(2004)}]{Pettini2004}
{Pettini}, M., \& {Pagel}, B. E.~J. 2004, \mnras, 348, L59,
  \dodoi{10.1111/j.1365-2966.2004.07591.x}

\bibitem[{{Phillips} {et~al.}(2013){Phillips}, {Simon}, {Morrell}, {Burns},
  {Cox}, {Foley}, {Karakas}, {Patat}, {Sternberg}, {Williams}, {Gal-Yam},
  {Hsiao}, {Leonard}, {Persson}, {Stritzinger}, {Thompson}, {Campillay},
  {Contreras}, {Folatelli}, {Freedman}, {Hamuy}, {Roth}, {Shields}, {Suntzeff},
  {Chomiuk}, {Ivans}, {Madore}, {Penprase}, {Perley}, {Pignata}, {Preston}, \&
  {Soderberg}}]{Phillips2013}
{Phillips}, M.~M., {Simon}, J.~D., {Morrell}, N., {et~al.} 2013, \apj, 779, 38,
  \dodoi{10.1088/0004-637X/779/1/38}

\bibitem[{{Piro} {et~al.}(2017){Piro}, {Muhleisen}, {Arcavi}, {Sand },
  {Tartaglia}, \& {Valenti}}]{Piro17}
{Piro}, A.~L., {Muhleisen}, M., {Arcavi}, I., {et~al.} 2017, \apj, 846, 94,
  \dodoi{10.3847/1538-4357/aa8595}

\bibitem[{{Poznanski} {et~al.}(2010){Poznanski}, {Nugent}, \&
  {Filippenko}}]{Poznanski2010}
{Poznanski}, D., {Nugent}, P.~E., \& {Filippenko}, A.~V. 2010, \apj, 721, 956,
  \dodoi{10.1088/0004-637X/721/2/956}

\bibitem[{{Poznanski} {et~al.}(2012){Poznanski}, {Prochaska}, \&
  {Bloom}}]{Poznanski2012}
{Poznanski}, D., {Prochaska}, J.~X., \& {Bloom}, J.~S. 2012, \mnras, 426, 1465,
  \dodoi{10.1111/j.1365-2966.2012.21796.x}

\bibitem[{{Quataert} \& {Shiode}(2012)}]{Quataert2012}
{Quataert}, E., \& {Shiode}, J. 2012, \mnras, 423, L92,
  \dodoi{10.1111/j.1745-3933.2012.01264.x}

\bibitem[{{Rabinak} \& {Waxman}(2011)}]{Rabinak2011}
{Rabinak}, I., \& {Waxman}, E. 2011, \apj, 728, 63,
  \dodoi{10.1088/0004-637X/728/1/63}

\bibitem[{{Reichart} {et~al.}(2005){Reichart}, {Nysewander}, {Moran},
  {Bartelme}, {Bayliss}, {Foster}, {Clemens}, {Price}, {Evans}, {Salmonson},
  {Trammell}, {Carney}, {Keohane}, \& {Gotwals}}]{Reichart2005}
{Reichart}, D., {Nysewander}, M., {Moran}, J., {et~al.} 2005, Nuovo Cimento C
  Geophysics Space Physics C, 28, 767, \dodoi{10.1393/ncc/i2005-10149-6}

\bibitem[{{Rui} {et~al.}(2019){Rui}, {Wang}, {Mo}, {Xiang}, {Zhang}, {Maund},
  {Gal-Yam}, {Wang}, \& {Zhang}}]{Rui2019}
{Rui}, L., {Wang}, X., {Mo}, J., {et~al.} 2019, \mnras, 485, 1990,
  \dodoi{10.1093/mnras/stz503}

\bibitem[{{Sapir} \& {Waxman}(2017)}]{Sapir2017}
{Sapir}, N., \& {Waxman}, E. 2017, \apj, 838, 130,
  \dodoi{10.3847/1538-4357/aa64df}

\bibitem[{{Savitzky} \& {Golay}(1964)}]{Savitzky1964}
{Savitzky}, A., \& {Golay}, M.~J.~E. 1964, Analytical Chemistry, 36, 1627

\bibitem[{{Schlafly} \& {Finkbeiner}(2011)}]{Schlafly2011}
{Schlafly}, E.~F., \& {Finkbeiner}, D.~P. 2011, \apj, 737, 103,
  \dodoi{10.1088/0004-637X/737/2/103}

\bibitem[{{Schmelling}(1995)}]{Schmelling1995}
{Schmelling}, M. 1995, \physscr, 51, 676, \dodoi{10.1088/0031-8949/51/6/002}

\bibitem[{{Shappee} {et~al.}(2014){Shappee}, {Prieto}, {Grupe}, {Kochanek},
  {Stanek}, {De Rosa}, {Mathur}, {Zu}, {Peterson}, {Pogge}, {Komossa}, {Im},
  {Jencson}, {Holoien}, {Basu}, {Beacom}, {Szczygie{\l}}, {Brimacombe},
  {Adams}, {Campillay}, {Choi}, {Contreras}, {Dietrich}, {Dubberley},
  {Elphick}, {Foale}, {Giustini}, {Gonzalez}, {Hawkins}, {Howell}, {Hsiao},
  {Koss}, {Leighly}, {Morrell}, {Mudd}, {Mullins}, {Nugent}, {Parrent},
  {Phillips}, {Pojmanski}, {Rosing}, {Ross}, {Sand}, {Terndrup}, {Valenti},
  {Walker}, \& {Yoon}}]{Shappee14}
{Shappee}, B.~J., {Prieto}, J.~L., {Grupe}, D., {et~al.} 2014, \apj, 788, 48,
  \dodoi{10.1088/0004-637X/788/1/48}

\bibitem[{{Simcoe} {et~al.}(2013){Simcoe}, {Burgasser}, {Schechter}, {Fishner},
  {Bernstein}, {Bigelow}, {Pipher}, {Forrest}, {McMurtry}, {Smith}, \&
  {Bochanski}}]{Simcoe2013}
{Simcoe}, R.~A., {Burgasser}, A.~J., {Schechter}, P.~L., {et~al.} 2013, \pasp,
  125, 270, \dodoi{10.1086/670241}

\bibitem[{{Smartt}(2015)}]{Smartt2015}
{Smartt}, S.~J. 2015, \pasa, 32, e016, \dodoi{10.1017/pasa.2015.17}

\bibitem[{{Smartt} {et~al.}(2009){Smartt}, {Eldridge}, {Crockett}, \&
  {Maund}}]{Smartt09}
{Smartt}, S.~J., {Eldridge}, J.~J., {Crockett}, R.~M., \& {Maund}, J.~R. 2009,
  \mnras, 395, 1409, \dodoi{10.1111/j.1365-2966.2009.14506.x}

\bibitem[{{Smith} {et~al.}(2020){Smith}, {Smartt}, {Young}, {Tonry}, {Denneau},
  {Flewelling}, {Heinze}, {Weiland}, {Stalder}, {Rest}, {Stubbs}, {Anderson},
  {Chen}, {Clark}, {Do}, {F{\"o}rster}, {Fulton}, {Gillanders}, {McBrien},
  {O'Neill}, {Srivastav}, \& {Wright}}]{ATLAS2}
{Smith}, K.~W., {Smartt}, S.~J., {Young}, D.~R., {et~al.} 2020, arXiv e-prints,
  arXiv:2003.09052.
\newblock \doarXiv{2003.09052}

\bibitem[{{Smith}(2014)}]{Smith2014}
{Smith}, N. 2014, \araa, 52, 487, \dodoi{10.1146/annurev-astro-081913-040025}

\bibitem[{{Smith} \& {Arnett}(2014)}]{Smith2014a}
{Smith}, N., \& {Arnett}, W.~D. 2014, \apj, 785, 82,
  \dodoi{10.1088/0004-637X/785/2/82}

\bibitem[{{Spiro} {et~al.}(2014){Spiro}, {Pastorello}, {Pumo}, {Zampieri},
  {Turatto}, {Smartt}, {Benetti}, {Cappellaro}, {Valenti}, {Agnoletto},
  {Altavilla}, {Aoki}, {Brocato}, {Corsini}, {Di Cianno}, {Elias-Rosa},
  {Hamuy}, {Enya}, {Fiaschi}, {Folatelli}, {Desidera}, {Harutyunyan}, {Howell},
  {Kawka}, {Kobayashi}, {Leibundgut}, {Minezaki}, {Navasardyan}, {Nomoto},
  {Mattila}, {Pietrinferni}, {Pignata}, {Raimondo}, {Salvo}, {Schmidt},
  {Sollerman}, {Spyromilio}, {Taubenberger}, {Valentini}, {Vennes}, \&
  {Yoshii}}]{Spiro2014}
{Spiro}, S., {Pastorello}, A., {Pumo}, M.~L., {et~al.} 2014, \mnras, 439, 2873,
  \dodoi{10.1093/mnras/stu156}

\bibitem[{{Sukhbold} {et~al.}(2016){Sukhbold}, {Ertl}, {Woosley}, {Brown}, \&
  {Janka}}]{Sukhbold2016}
{Sukhbold}, T., {Ertl}, T., {Woosley}, S.~E., {Brown}, J.~M., \& {Janka}, H.~T.
  2016, \apj, 821, 38, \dodoi{10.3847/0004-637X/821/1/38}

\bibitem[{{Tak{\'a}ts} {et~al.}(2015){Tak{\'a}ts}, {Pignata}, {Pumo},
  {Paillas}, {Zampieri}, {Elias-Rosa}, {Benetti}, {Bufano}, {Cappellaro},
  {Ergon}, {Fraser}, {Hamuy}, {Inserra}, {Kankare}, {Smartt}, {Stritzinger},
  {Van Dyk}, {Haislip}, {LaCluyze}, {Moore}, \& {Reichart}}]{Takats2015}
{Tak{\'a}ts}, K., {Pignata}, G., {Pumo}, M.~L., {et~al.} 2015, \mnras, 450,
  3137, \dodoi{10.1093/mnras/stv857}

\bibitem[{{Tartaglia} {et~al.}(2018){Tartaglia}, {Sand}, {Valenti}, {Wyatt},
  {Anderson}, {Arcavi}, {Ashall}, {Botticella}, {Cartier}, {Chen}, {Cikota},
  {Coulter}, {Della Valle}, {Foley}, {Gal-Yam}, {Galbany}, {Gall}, {Haislip},
  {Harmanen}, {Hosseinzadeh}, {Howell}, {Hsiao}, {Inserra}, {Jha}, {Kankare},
  {Kilpatrick}, {Kouprianov}, {Kuncarayakti}, {Maccarone}, {Maguire},
  {Mattila}, {Mazzali}, {McCully}, {Meland ri}, {Morrell}, {Phillips},
  {Pignata}, {Piro}, {Prentice}, {Reichart}, {Rojas-Bravo}, {Smartt}, {Smith},
  {Sollerman}, {Stritzinger}, {Sullivan}, {Taddia}, \& {Young}}]{Tartaglia2018}
{Tartaglia}, L., {Sand}, D.~J., {Valenti}, S., {et~al.} 2018, \apj, 853, 62,
  \dodoi{10.3847/1538-4357/aaa014}

\bibitem[{{Tomasella} {et~al.}(2013){Tomasella}, {Cappellaro}, {Fraser},
  {Pumo}, {Pastorello}, {Pignata}, {Benetti}, {Bufano}, {Dennefeld},
  {Harutyunyan}, {Iijima}, {Jerkstrand }, {Kankare}, {Kotak}, {Magill},
  {Nascimbeni}, {Ochner}, {Siviero}, {Smartt}, {Sollerman}, {Stanishev},
  {Taddia}, {Taubenberger}, {Turatto}, {Valenti}, {Wright}, \&
  {Zampieri}}]{Tomasella2013}
{Tomasella}, L., {Cappellaro}, E., {Fraser}, M., {et~al.} 2013, \mnras, 434,
  1636, \dodoi{10.1093/mnras/stt1130}

\bibitem[{{Tonry}(2011)}]{Tonry2011}
{Tonry}, J.~L. 2011, \pasp, 123, 58, \dodoi{10.1086/657997}

\bibitem[{{Utrobin} \& {Chugai}(2015)}]{Utrobin2015}
{Utrobin}, V.~P., \& {Chugai}, N.~N. 2015, \aap, 575, A100,
  \dodoi{10.1051/0004-6361/201424822}

\bibitem[{{Utrobin} \& {Chugai}(2017)}]{Utrobin2017}
---. 2017, \mnras, 472, 5004, \dodoi{10.1093/mnras/stx2415}

\bibitem[{{Valenti} {et~al.}(2018){Valenti}, {Sand}, \& {Wyatt}}]{Valenti2018}
{Valenti}, S., {Sand}, D.~J., \& {Wyatt}, S. 2018, Transient Name Server
  Discovery Report, 2018-876, 1

\bibitem[{{Valenti} {et~al.}(2008){Valenti}, {Benetti}, {Cappellaro}, {Patat},
  {Mazzali}, {Turatto}, {Hurley}, {Maeda}, {Gal-Yam}, {Foley}, {Filippenko},
  {Pastorello}, {Challis}, {Frontera}, {Harutyunyan}, {Iye}, {Kawabata},
  {Kirshner}, {Li}, {Lipkin}, {Matheson}, {Nomoto}, {Ofek}, {Ohyama}, {Pian},
  {Poznanski}, {Salvo}, {Sauer}, {Schmidt}, {Soderberg}, \&
  {Zampieri}}]{Valenti2008}
{Valenti}, S., {Benetti}, S., {Cappellaro}, E., {et~al.} 2008, \mnras, 383,
  1485, \dodoi{10.1111/j.1365-2966.2007.12647.x}

\bibitem[{{Valenti} {et~al.}(2014){Valenti}, {Sand}, {Pastorello}, {Graham},
  {Howell}, {Parrent}, {Tomasella}, {Ochner}, {Fraser}, {Benetti}, {Yuan},
  {Smartt}, {Maund}, {Arcavi}, {Gal-Yam}, {Inserra}, \& {Young}}]{Valenti2014}
{Valenti}, S., {Sand}, D., {Pastorello}, A., {et~al.} 2014, \mnras, 438, L101,
  \dodoi{10.1093/mnrasl/slt171}

\bibitem[{{Valenti} {et~al.}(2016){Valenti}, {Howell}, {Stritzinger}, {Graham},
  {Hosseinzadeh}, {Arcavi}, {Bildsten}, {Jerkstrand}, {McCully}, {Pastorello},
  {Piro}, {Sand}, {Smartt}, {Terreran}, {Baltay}, {Benetti}, {Brown},
  {Filippenko}, {Fraser}, {Rabinowitz}, {Sullivan}, \& {Yuan}}]{Valenti2016}
{Valenti}, S., {Howell}, D.~A., {Stritzinger}, M.~D., {et~al.} 2016, \mnras,
  459, 3939, \dodoi{10.1093/mnras/stw870}

\bibitem[{{Van Dyk} {et~al.}(2003){Van Dyk}, {Li}, \& {Filippenko}}]{VanDyk03}
{Van Dyk}, S.~D., {Li}, W., \& {Filippenko}, A.~V. 2003, \pasp, 115, 1,
  \dodoi{10.1086/345748}

\bibitem[{{Walmswell} \& {Eldridge}(2012)}]{Walmswell2012}
{Walmswell}, J.~J., \& {Eldridge}, J.~J. 2012, \mnras, 419, 2054,
  \dodoi{10.1111/j.1365-2966.2011.19860.x}

\bibitem[{{W}es {M}c{K}inney(2010)}]{mckinney-proc-scipy-2010}
{W}es {M}c{K}inney. 2010, in {P}roceedings of the 9th {P}ython in {S}cience
  {C}onference, ed. {S}t\'efan van~der {W}alt \& {J}arrod {M}illman, 56 -- 61,
  \dodoi{10.25080/Majora-92bf1922-00a}

\bibitem[{{Willick} {et~al.}(1997){Willick}, {Courteau}, {Faber}, {Burstein},
  {Dekel}, \& {Strauss}}]{Willick1997}
{Willick}, J.~A., {Courteau}, S., {Faber}, S.~M., {et~al.} 1997, \apjs, 109,
  333, \dodoi{10.1086/312983}

\bibitem[{{Woosley} \& {Heger}(2007)}]{Woosley2007}
{Woosley}, S.~E., \& {Heger}, A. 2007, \physrep, 442, 269,
  \dodoi{10.1016/j.physrep.2007.02.009}

\bibitem[{{Yang} {et~al.}(2017){Yang}, {Valenti}, {Cappellaro}, {Sand },
  {Tartaglia}, {Corsi}, {Reichart}, {Haislip}, \& {Kouprianov}}]{Yang17}
{Yang}, S., {Valenti}, S., {Cappellaro}, E., {et~al.} 2017, \apjl, 851, L48,
  \dodoi{10.3847/2041-8213/aaa07d}

\bibitem[{{Yang} {et~al.}(2019){Yang}, {Sand}, {Valenti}, {Cappellaro},
  {Tartaglia}, {Wyatt}, {Corsi}, {Reichart}, {Haislip}, {Kouprianov}, \&
  {(DLT40 Collaboration}}]{Yang19}
{Yang}, S., {Sand}, D.~J., {Valenti}, S., {et~al.} 2019, \apj, 875, 59,
  \dodoi{10.3847/1538-4357/ab0e06}

\bibitem[{{Yaron} \& {Gal-Yam}(2012)}]{Yaron2012}
{Yaron}, O., \& {Gal-Yam}, A. 2012, \pasp, 124, 668, \dodoi{10.1086/666656}

\bibitem[{{Yaron} {et~al.}(2017){Yaron}, {Perley}, {Gal-Yam}, {Groh}, {Horesh},
  {Ofek}, {Kulkarni}, {Sollerman}, {Fransson}, {Rubin}, {Szabo}, {Sapir},
  {Taddia}, {Cenko}, {Valenti}, {Arcavi}, {Howell}, {Kasliwal}, {Vreeswijk},
  {Khazov}, {Fox}, {Cao}, {Gnat}, {Kelly}, {Nugent}, {Filippenko}, {Laher},
  {Wozniak}, {Lee}, {Rebbapragada}, {Maguire}, {Sullivan}, \&
  {Soumagnac}}]{Yaron2017}
{Yaron}, O., {Perley}, D.~A., {Gal-Yam}, A., {et~al.} 2017, Nature Physics, 13,
  510, \dodoi{10.1038/nphys4025}

\end{thebibliography}
\bibliographystyle{aasjournal} 

\appendix

\begin{deluxetable*}{ccccccc}
\tablenum{A1}
\tablecaption{SN 2018cuf Optical Photometry\label{tab:photometry}}
\tablewidth{0pt}
\tablehead{
\colhead{Date} & \colhead{Julian Date (Days)} & \colhead{Phase (Days)} &
\colhead{Magnitude} & \colhead{Magnitude Error} & \colhead{Filter} &
\colhead{Source}}
\startdata
2018-06-19&2458288.78&-4.03&$>$19.70&0.0&Open&Prompt5\\
2018-06-21&2458290.83&-1.98&$>$19.68&0.0&Open&Prompt5\\
2018-06-22&2458291.74&-1.07&$>$19.38&0.0&Open&Prompt5\\
2018-06-23&2458292.86&0.05&17.44&0.02&Open&Prompt5\\
2018-06-23&2458293.20&0.39&16.63&0.02&U&COJ 1m\\
2018-06-23&2458293.21&0.40&16.68&0.02&U&COJ 1m\\
2018-06-23&2458293.21&0.40&17.34&0.02&B&COJ 1m\\
2018-06-23&2458293.21&0.41&17.34&0.02&B&COJ 1m\\
2018-06-23&2458293.22&0.41&17.35&0.01&V&COJ 1m\\
2018-06-23&2458293.22&0.41&17.34&0.01&V&COJ 1m\\
2018-06-23&2458293.22&0.41&17.21&0.01&g&COJ 1m\\
\enddata{}
\tablecomments{This table will be published in its entirety in a machine-readable format.  A portion is shown here for guidance regarding its form and content.}
\end{deluxetable*}

\begin{deluxetable*}{ccccccc}
\tablenum{A2}
\tablecaption{SN 2018cuf Spectra\label{tab:spectra}}
\tablewidth{0pt}
\tablehead{
\colhead{UT Date} & \colhead{Julian Date (Days)} & \colhead{Phase (Days)} &
\colhead{Telescope} & \colhead{Instrument} & \colhead{Resolution ($\lambda/\Delta\lambda$)} & wavelenth range ($\rm \AA$)}
\startdata 
2018-06-24&2458293.73&0.92&Gemini&GMOS&1688&3916-7069\\
2018-06-24&2458294.50&1.69&SALT&RSS&600-2000&3533-7449\\
2018-06-25&2458294.50&1.69&SALT&RSS&600-2000&3497-7431\\
2018-06-27&2458296.50&3.69&Gemini&F2&900&9853-18081\\
2018-06-28&2458298.49&5.68&SALT&RSS&600-2000&3495-9396\\
2018-07-01&2458301.05&8.24&FTN&FLOYDS&400-700&4796-8996\\
2018-07-03&2458303.11&10.30&FTN&FLOYDS&400-700&3498-9999\\
2018-07-04&2458303.56&10.75&SALT&RSS&600-2000&3494-9393\\
2018-07-06&2458305.58&12.77&SALT&RSS&600-2000&3495-9396\\
2018-07-08&2458307.50&14.69&Gemini&F2&900&9851-18082\\
2018-07-10&2458310.07&17.26&FTN&FLOYDS&400-700&3498-9998\\
2018-07-11&2458310.50&17.69&Gemini&F2&900&9851-18081\\
2018-07-12&2458311.50&18.69&Magellan&MIKE&40000&4832-9415\\
2018-07-14&2458314.00&21.19&FTN&FLOYDS&400-700&3497-9998\\
2018-07-18&2458318.29&25.48&FTN&FLOYDS&400-700&3497-9999\\
2018-07-26&2458326.38&33.57&SALT&RSS&600-2000&3497-9398\\
2018-07-31&2458331.16&38.35&FTN&FLOYDS&400-700&4796-9996\\
2018-08-01&2458331.50&38.69&Gemini&F2&900&9847-18080\\
2018-08-09&2458339.96&47.15&FTN&FLOYDS&400-700&3498-9997\\
2018-08-18&2458348.92&56.11&FTN&FLOYDS&400-700&3497-9997\\
2018-08-20&2458350.50&57.69&Magellan&FIRE&300-500&7700-25269\\
2018-09-07&2458368.95&76.14&FTN&FLOYDS&400-700&3497-9997\\
2018-09-12&2458373.50&80.69&Magellan&FIRE&300-500&7755-25277\\
2018-09-16&2458377.92&85.11&FTN&FLOYDS&400-700&3498-9998\\
2018-09-18&2458379.96&87.15&FTN&FLOYDS&400-700&3497-9997\\
2018-09-25&2458386.50&93.69&Magellan&FIRE&300-500&7762-25297\\
2018-10-06&2458398.27&105.46&SALT&RSS&600-2000&3496-9395\\
2018-10-07&2458398.96&106.15&FTN&FLOYDS&400-700&3498-9998\\
2018-10-25&2458417.34&124.53&SALT&RSS&600-2000&3495-9394\\
2018-12-14&2458466.53&173.72&VLT&FORS&500&4608-8645\\
2019-04-12&2458585.50&292.69&VLT&MUSE&1700-3500&4750-9351\\
2019-05-25&2458628.65&335.84&SALT&RSS&600-2000&3898-8719\\
2019-05-26&2458629.53&336.72&SALT&RSS&600-2000&5899-8870\\
2019-06-23&2458658.46&365.65&SALT&RSS&600-2000&5900-9002\\
2019-10-19&2458776.29&483.48&SALT&RSS&600-2000&3921-7798\\
\enddata{}
\end{deluxetable*}


\end{CJK*}
\end{document}